\documentclass[english,aps,nofootinbib,twocolumn,preprintnumbers,nobalancelastpage,floatfix,showpacs]{revtex4-1}

\usepackage[latin1]{inputenc}
\pdfoutput = 1
\usepackage{bm}
\usepackage{color}
\usepackage{graphicx}
\usepackage{cancel}
\usepackage{amsfonts}
\usepackage{amssymb}
\usepackage{amsmath}
\usepackage{calc}
\usepackage{dcolumn}
\usepackage{mathrsfs}

\newcommand{\AddrRWTH}{%
Institut f\"ur Theoretische Teilchenphysik und Kosmologie, RWTH Aachen, 52056 Aachen, Germany}
\newcommand{\AddrGRAPPA}{%
GRAPPA Institute, University of Amsterdam, Science Park 904, 1090 GL Amsterdam, Netherlands}

\begin{document}

\title{Chasing a consistent picture for dark matter direct searches}

\author{Chiara~Arina}
\affiliation{\AddrGRAPPA}
\affiliation{\AddrRWTH}

\begin{abstract}
In this paper we assess the present status of dark matter direct searches by means of Bayesian statistics. We consider three particle physics models for spin-independent dark matter interaction with nuclei: elastic, inelastic and isospin violating scattering. We shortly present the state of the art for the three models, marginalising over experimental systematics and astrophysical uncertainties. Whatever the scenario is, XENON100 appears to challenge the detection region of DAMA, CoGeNT and CRESST. The first aim of this study is to rigorously quantify the significance of the inconsistency between XENON100 data and the combined set of detection (DAMA, CoGeNT and CRESST together), performing two statistical tests based on the Bayesian evidence. We show that XENON100 and the combined set are inconsistent at least at $2\sigma$ level in all scenarios but inelastic scattering, for which the disagreement drops to $1\sigma$ level. Secondly we consider only the combined set and hunt the best particle physics model that accounts for the events, using Bayesian model comparison. The outcome between elastic and isospin violating scattering is inconclusive, with the odds $2:1$, while inelastic scattering is disfavoured with the odds of $1:32$ because of CoGeNT data. Our results are robust under reasonable prior assumptions. We conclude that the simple elastic scattering remains the best model to explain the detection regions, since the data do not support extra free parameters. The outcome of consistency tests implies that either a better understanding of astrophysical and experimental uncertainties is needed and the strength of belief in certain data sets should be revised, either the dark matter theoretical model is at odds with the data.
 \end{abstract}
\pacs{95.35.+d, 95.30.Cq}
\maketitle

\section{Introduction}\label{sec:intro}

The last years have seen an intense activity in direct searches for dark matter (DM) candidates, in particular weakly interacting massive particles (WIMPs). Three experiments support a hint of detection in the low DM mass regime: CoGeNT~\cite{Aalseth:2011wp}, with an excess that follows a modulated behavior, CRESST-II~\cite{Angloher:2011uu} (CRESST from now on) with 67 events that can not be fully accounted for by known backgrounds, besides the thirteen-years signal at DAMA/LIBRA~\cite{Bernabei:2010mq} (hereafter DAMA), which shows an annual modulation compatible with WIMP predictions. Alongside these `signals', stands the series of null result experiments, most notably XENON100~\cite{:2012nq} (Xe100 henceforth), which has the world strongest exclusion limit at present. The (in)compatibility between the low mass hints themselves and the several upper limits has been discussed in a variety of papers, see {\it e.g.}~\cite{Fornengo:2011sz,DelNobile:2012tx,Frandsen:2011gi,Kopp:2011yr,*Feng:2008dz,*Gao:2011ka,*Kelso:2011gd,*Hooper:2012ft,*Foot:2012rk,*Bergstrom:2012fi,*Hooper:2012cw,*Cline:2012ei,*Perelstein:2012qg,*Jin:2012jn} for recent analyses in both model independent and specific DM scenarios. In this study our purpose is to use the tools of Bayesian statistics to investigate quantitatively the tension between experiments and to find which particle physics model provides the best compromise for the low mass hints, motivated by the very recent data release of Xe100 and the fact that an excess is likely still present in the new science run of CoGeNT~\cite{collarIDM}. 

Before heading towards the main intent of the paper, however, we wish to extend the Bayesian analysis pursued in~\cite{Arina:2011si} to the most recent experimental results and to distinct particle physics interactions. We employ the same procedure as in~\cite{Arina:2011si} to include experimental systematics in the likelihood and to encompass astrophysical uncertainties using a motivated DM density profile with the related velocity distribution. The inclusion of astrophysical uncertainties is becoming a common procedure, starting from~\cite{Ullio:2000bf,*Strigari:2009zb,*Belli:2011kw} for analysis of experimental results to~\cite{Pato:2010zk,*Strege:2012kv,Kavanagh:2012nr,*Fairbairn:2012zs} for reconstruction of WIMP parameters and forecasts. We consider, in addition to CoGeNT, DAMA and Xe100, the CRESST excess and KIMS~\cite{Kim:2012rz} experiment. It is worth to analyse as well the exclusion bounds released by bubble chamber experiments, like PICASSO~\cite{Archambault:2012pm} and SIMPLE-II~\cite{Felizardo:2011uw}. These experiments start to have a total exposure sensitive to the cross-sections questioned by the low mass hints. Several scenarios of particle physics other than elastic spin-independent interaction have been proposed, trying to accommodate the exclusion bounds and DAMA, CoGeNT, CRESST excesses: {\it e.g.} inelastic DM~\cite{TuckerSmith:2001hy}, isospin violating scattering~\cite{Feng:2011vu,Frandsen:2011ts}, long range forces~\cite{Fornengo:2011sz,DelNobile:2012tx} or composite DM~\cite{Khlopov:2008ty}. Here, we consider the class of spin-independent interaction, namely elastic, inelastic and isospin violating scattering. These are nested models: the more complicated models ({\it e.g.} with additional degrees of freedom) can be reduced to the simplest one by fixing at a certain value the extra free parameters. We present inference for all the experiments listed above to establish the state of the art of current DM direct detection in each particle physics model considered, having marginalised over all nuisance parameters. This will be the ground for our Bayesian analysis, explained in the following. 

The outcome of parameter inference signals a disagreement between the detection regions and the exclusion bounds `by eye': every experiment is evaluated separately and then all the contours are displayed together in a single plot, showing marginal or no overlap. Firstly we feel that it would be interesting to further investigate this tension and to make use of statistical tools to quantify the degree of inconsistency between Xe100 exclusion bound and DAMA, CoGeNT and CRESST together (to which we will refer hereafter as combined set), in the nested model framework described above. Our purpose is to re-consider the problem of the tension between all these experimental results with two statistical tests: the predictive likelihood ratio, or $\mathscr{L}$-test, and the $\mathscr{R}$-test, after~\cite{Feroz:2009dv} and introduced below. Both tests are based on the Bayesian evidence~\cite{Trotta:2008qt,Sellke:2001}, which is by definition the likelihood averaged over parameter space weighted by the prior probability of the parameters. These tests are therefore performed in data space. For a given set of data (Xe100 + combined set), we allow ourself to change the outcome of a subset of it (Xe100 data), keeping the rest fixed (combined set), to check wether a different observed value would improuve or diminish the agreement between the whole set. The result of each test will provide the statistical significance of the (dis)agreement between Xe100 data and the detection regions in every particle physics scenario.

Since the tests will point out to an incompatibility between Xe100 and the detection regions, it does not make sense to combine all those experiments together. In the second part of this study, we then consider only the detection regions and apply Bayesian model comparison to select which one of the nested particle physics models explains better the observations. 
Indeed, a scientific question that might be asked is about the probability of competing models under the data. This question can be assessed in the framework of Bayesian model comparison, by means of the Bayesian evidence, which automatically incorporates the notion of Occam's razor. Indeed models that properly fit the data are rewarded through a favourable likelihood function, while models that are unpredictive are penalised by the larger parameter volume over which the likelihood must be averaged. The use of Bayesian model comparison is not so common in particle physics, however see {\it e.g.}~\cite{Feroz:2009dv,Cabrera:2010xx,Arina:2011xu,Arina:2011zh}. 

The paper is organised as follows. In section~\ref{sec:stat} we define the statistical framework for Bayesian inference, model comparison and consistency checks. The succeeding section~\ref{sec:dd} resumes in short the main feature of direct detection rates and defines the particle physics models we wish to compare. In section~\ref{sec:expdlkh} we briefly define the likelihood for each experiment we consider and include the astrophysical uncertainties, while the details are given in appendix~\ref{sec:appa}. The up-to-date situation for DM direct searches is described in~\ref{sec:soa} (with more details in appendix~\ref{sec:appb}), and we present the outcome for Bayesian tests and model comparison in section~\ref{sec:res}. Our conclusions are summarised in section~\ref{sec:concl}.

\section{Setup of the statistical framework}\label{sec:stat}
\subsection{Parameter inference}
Given a set of parameters $\theta$ defining a model ${\mathcal M}$, we are interested to compute their posterior probability distribution function (pdf) $p(\theta | d, {\mathcal M})$ 
via Bayes' theorem, namely
\begin{equation}\label{eq:BT}
p(\theta | d, {\mathcal M}) = \frac{{\mathcal L}(\theta | {\mathcal M}) p(\theta|{\mathcal M})}{p(d|{\mathcal M})}.
\end{equation}
Here, $d$ are the data under consideration, ${\mathcal L}(\theta | {\mathcal M}) \equiv  p(d|\theta,{\mathcal M})$ the likelihood function,
and $p(\theta|{\mathcal M})$ is the prior pdf for the parameters under the model. The quantity $p(d|{\mathcal M})$, defined as
\begin{equation}
\label{eq:evidence}
{p}(d|\mathcal{M}) \equiv \int \mathcal{L}(\theta| \mathcal{M})p(\theta|\mathcal{M}) {\rm d} \theta\,,
\end{equation}
is called the Bayesian evidence.

The posterior pdf contains all the necessary information for the interpretation of  the data, however typically its dimensionality is reduced to $n=1,2$ by integrating out the $m$ nuisance parameter directions $\psi$ for `graphical' purposes, yielding to the so-called marginal posterior pdf
\begin{eqnarray}
 \label{eq:marg}
& \mathcal{P}_{\rm mar}(\theta_1,...,\theta_n | d)  \propto\nonumber\\
& \int {\rm d}\psi_1
 ... {\rm d}\psi_m \ {\cal P}( \theta_1,...,\theta_n,\psi_1,...,\psi_m|d) \,, 
\end{eqnarray} 
which is used to construct constraints on the remaining parameters as well.

Provided the data are sufficiently constraining the marginal posterior is usually insensitive to the choice of prior. For data that can only provide an upper or a lower bound on a parameter however, the properties of the inferred posterior and the boundaries of credible regions can vary significantly with the choice of prior as well as its limits $\theta_{\rm min}$ and $\theta_{\rm max}$, making an objective interpretation of the
results rather difficult. This is the case of exclusion limits: for them instead of computing credible intervals from 
the fractional volume of the marginal posterior we construct intervals based on the volume of the marginal posterior in $S$-space, where $S$ is the expected WIMP signal, using a uniform prior on $S$ with a lower boundary at zero~\cite{Helene:1982pb}. To distinguish these $S$-based credible intervals from the conventional ones based on the volume of the marginal posterior pdf, we label them with a subscript $S$, {\it e.g.} $90_S\%$. For more details on this construction we refer to~\cite{Arina:2011si}.

\subsection{Model comparison}

Bayesian inference is based on the posterior pdf for the parameters $\theta$, and it assumes that the model under consideration, $\mathcal M$, is the correct one. We can however expand the inferential framework to the viability of the model itself or of the relative performance of alternative possible models as explanation for the data. The formalism of Bayesian model comparison automatically balances the quality of the model's  fit to the data against its predictiveness, that is the best model achieves the optimum compromise between quality of fit and predictiveness and will have the highest posterior probability. In this sense, the methodology of  Bayesian model selection can be interpreted as  a quantitative expression of the Occam's razor principle of simplicity. The Bayesian evidence takes into account the entire allowed range of parameters and it incorporates a well defined notion of probability for a model against another one. We define here the basics, while for a more in-depth discussion see {\it e.g.}~\cite{Kunz:2006mc,Trotta:2008qt}. 

From equation~\ref{eq:BT}, the posterior odds between two competing models $\mathcal{M}_0$ and $\mathcal{M}_1$ are given by
\begin{equation}
\frac{p(\mathcal{M}_1|d)}{p(\mathcal{M}_0|d)} = B\,  \frac{p(\mathcal{M}_1) }{p(\mathcal{M}_0) }\,,
\end{equation}
where  
\begin{equation}
\label{eq:bayesfactor}
B \equiv \frac{p(d|\mathcal{M}_1)}{p(d|\mathcal{M}_0)}\,
\end{equation}
is the Bayes factor, defined as the ratio of the models' evidences. The Bayes factor $B$ represents an update from our prior belief in the odds of two competing models $p(\mathcal{M}_1)/p(\mathcal{M}_0)$
to the posterior odds $p(\mathcal{M}_1|d)/p(\mathcal{M}_0|d)$.  If the two models have non-committal prior ($p(\mathcal{M}_1)=p(\mathcal{M}_0)$) the Bayes factor alone determines the outcome of the model comparison. Considering the logarithm of the Bayes factor, a positive value means that the model $\mathcal{M}_1$ is preferred over the model $\mathcal{M}_0$ as a description of the 
experimental data, and vice versa. The correspondence between the actual value of the Bayes factor and strength of belief follows the convention set down by Jeffreys' scale shown in table~\ref{tab:jef}.

From the definition of the Bayesian evidence in equation~(\ref{eq:evidence}), note how this quantity incorporates the notion of Occam's razor and penalises those models with excessive complexity unsupported by the data for wasted parameter space. Increasing the dimensionality of the parameter space without significantly enhancing the likelihood $\mathcal{L}(d|\theta, \mathcal{M})$ in the new parameter directions reduces the evidence. Unpredictive priors $p(\theta|\mathcal{M})$, namely excessively broad compared with the width of the likelihood, dilute the evidence as well. Hence a sensitivity analysis of the results of Bayesian model selection is necessary, since the choice of priors is usually not unique. This analysis assesses the dependence of $\ln B$ on a reasonable change of priors as follows. If the models $\mathcal{M}_0$ and $\mathcal{M}_1$ are nested and their parameter priors separable, then the impact of changing the prior width on the Bayes factor may be estimated analytically using the Savage-Dickey density ratio (SDDR, see~\cite{Trotta:2005ar}). The SDDR ratio depends only on the prior of the extra parameter: indeed if the data are sufficiently constraining, the marginal posterior pdf will exhibit little dependence on the prior, therefore priors for common parameters factor out. If the prior of the extra parameter is a top-hat function, rescaling its width by a factor $\lambda$ will change $\ln B$ by approximately $-\ln \lambda$, as a consequence of priors being normalized to unity probability content~\cite{Arina:2011zh}.
 
For deciding whether the introduction of new parameters in the theory is necessary, the frequentist approach relies on the $\Delta \chi^2_{\rm eff}$, based on the evaluation of the likelihood at the best-fit point, and $p$-values, which return the probability of observing as extreme or more extreme values of the test statistic assuming the null hypothesis is true. For sake of reference we give as well the $\Delta\chi^2_{\rm eff}$ (defined as twice the difference between the best-fit likelihood values) and the classical $p$-values, following~\cite{Trotta:2008qt,Arina:2011zh}. For the nested models we consider, the extra parameters satisfy Chernoff's theorem~\cite{Chernoff:1954,*Shapiro:1988}, that is the null hypothesis sits on the boundary but the additional parameters are all defined under the null. The test statistics for the $p$-value is therefore a weighted sum of $\chi^2$ distributions.
\begin{table}[t]
\caption{Jeffreys' scale for grading the strength of evidence for two competing models $\mathcal{M}_0$ and $\mathcal{M}_1$, adapted from~\cite{Gordon:2007xm,Trotta:2008qt}.\label{tab:jef}}
\begin{center}
\begin{ruledtabular}
\begin{tabular}{lll}
$\ln B$ & Odds $\mathcal{M}_1: \mathcal{M}_0$ & Strength of evidence \\
\hline
$<-5.0$ & $< 1:150$ & Strong evidence for $\mathcal{M}_0$ \\
$-5.0 \to -2.5$  & $1:150 \to 1:12$ & Moderate evidence for $\mathcal{M}_0$ \\
$-2.5 \to -1.0$ & $1:12 \to 1:3$ & Weak evidence for $\mathcal{M}_0 $ \\
$-1.0 \to 1.0$ & $1:3 \to 3:1$ & Inconclusive\\
$1.0 \to 2.5$ & $3:1 \to 12:1$ & Weak evidence against $\mathcal{M}_0 $ \\
$2.5 \to 5.0$  & $12:1 \to 150:1$ & Moderate evidence against $\mathcal{M}_0$ \\
$> 5.0$ &$> 150:1$ & Strong evidence against $\mathcal{M}_0$ \\
\end{tabular}
\end{ruledtabular}
\end{center}
\end{table}

\subsection{$\mathscr{L}$-test and $\mathscr{R}$-test}

Model comparison is one application of Bayesian model selection, while another possibility is quantifying the consistency between two or more data sets (see {\it e.g.}~\cite{Feroz:2008wr,Feroz:2009dv,Cabrera:2010xx} for particle physics applications). Any obvious tension between experimental results is likely to be noticed by the `chi by eye', as it is common practice in direct detection analyses. Indeed outcomes from different experiments may push the model parameters to different corners of the parameter space. Here we claim that it is important to privilege a method that quantifies these discrepancies, as follows.

A full data set under consideration $d$ can be divided into two parts as $d=\{\mathscr{D},D\}$, where $\mathscr{D}$ is the subset we wish to test for compatibility with respect to the remaining data set $D$, which we take as reference.  The conditional evidence $p(\mathscr{D}|D)$ is the probability of measuring the data $\mathscr{D}$, knowing that the set $D$ has been measured:
\begin{equation}\label{eq:jointpdf}
p(\mathscr{D}|D) = \frac{p(\mathscr{D},D)}{p(D)}\,.
\end{equation}
Here $p(\mathscr{D},D)$ is the joint evidence,  that is the probability of measuring the whole set $d$ within the model under investigation. Note that this measure is independent on the actual values of the model parameters $\theta$, which have been integrated out by definition of evidence. Then $p(D)$ is the Bayesian evidence corresponding only to the data subset $D$ and is a normalization factor that will cancel out. The conditioning on the model $\mathcal{M}$ is understood in all the formulas of this section. We then define $\mathscr{D}^{\rm obs}$ as the observed value for the variable $\mathscr{D}$.

The first test we consider is called predictive likelihood test or $\mathscr{L}-$test. The consistency of $\mathscr{D}^{\rm obs}$ with the remaining data $D$ is evaluated by comparing $p(\mathscr{D}^{\rm obs}|D)$ with the value of $\mathscr{D}$ that maximises such probability, called $\mathscr{D}^{\rm max}$:
\begin{equation}\label{eq:ltest}
\mathscr{L}(\mathscr{D}^{\rm obs}|D) = \frac{p(\mathscr{D}^{\rm obs}|D)}{p(\mathscr{D}^{\rm max}|D)} = \frac{p(\mathscr{D}^{\rm obs},D)}{p(\mathscr{D}^{\rm max},D)}\,.
\end{equation}
The $\mathscr{L}$ distribution is simply given by the ratio of the joint evidences at the observed and maximal value, by means of equation~\ref{eq:jointpdf}. This is analogous to a likelihood ratio in data space, that is integrated over all possible values of the models' parameters. More precisely, we evaluate the joint evidence as a function of the possible outcome of the observation of the data set $\mathscr{D}$ while at the same time the set $D$ is kept fixed at its actual value. We take the freedom of varying the value of $\mathscr{D}$, assuming the same errors on systematics as reported by the experiment. Then we measure the relative probability of obtaining the observed data realization $\mathscr{D}^{\rm obs}$ to the maximum probability of the data set in question. If the outcome of the comparison, $\ln \mathscr{L}(\mathscr{D}^{\rm obs}|D)$, is close to zero both data sets are compatible with each other and with the model assumptions. If however $\ln \mathscr{L}(\mathscr{D}^{\rm obs}|D) \ll 0$ there is clearly a tension between $D$ and $\mathscr{D}$. This means that one should doubt the models' assumption or doubt $\mathscr{D}$ (or vice versa doubt the reference set) and look properly for systematics. The $\mathscr{L}$-test is weakly dependent on the prior choice, being a likelihood ratio by definition and can be evaluated on a significance scale alike $\Delta \chi^2$.

The second test we perform is the $\mathscr{R}$-test, called model comparison test as well. In this case we test two hypotheses, again in data space. Suppose that $\mathcal{H}_0$ states that all the data sets under scrutiny are compatible with each other and with the models' assumption. On the contrary $\mathcal{H}_1$ affirms that the observed experimental outcomes are inconsistent  so that each data set requires its own set of parameter values, since they privilege different regions in the parameter space. Then the Bayes factor between the two hypotheses, if we have no reason to prefer either $\mathcal{H}_0$ or $\mathcal{H}_1$, is given by
\begin{equation}\label{eq:rtest}
\mathscr{R}(\mathscr{D}^{\rm obs}) = \frac{p(\mathscr{D}^{\rm obs}, D | \mathcal{H}_0)}{p(\mathscr{D}^{\rm obs} | \mathcal{H}_1) p(D | \mathcal{H}_1)}\,.
\end{equation}
For positive value of $\ln \mathscr{R}(\mathscr{D}^{\rm obs})$ the data sets are compatible, while for negative values the alternative hypothesis $\mathcal{H}_1$ is preferred. The strength of evidence against/in favour of $\mathcal{H}_0$ is assessed by the Jeffreys' scale (table~\ref{tab:jef}) as for Bayesian model selection.
 
In this paper the data we wish to test by means of the $\mathscr{L}$-test is the number of observed events at Xe100 experiment, $\mathscr{D} \equiv N_{\rm events}$, while the reference data are given by the combined set $D=\{ \rm DAMA, CoGeNT, CRESST\}$. We investigate through the $\mathscr{R}$-test the hypothesis of compatibility between data sets: $\mathcal{H}_0$ believes that Xe100 outcome is consistent with the combined set, while $\mathcal{H}_1$ denotes the incompatibility hypothesis. 

The computation of the evidence $p(d|\mathcal{M})$ for each model $\mathcal{M}$ requires the evaluation of an integral over the parameter space. We use the ellipsoidal and multimodal nested-sampling algorithm implemented in the publicly available package \texttt{MultiNest} v2.12~\cite{Feroz:2007kg,Feroz:2008xx}.  We set $n_{\rm live} = 10000$, an efficiency factor of $10^{-4}$ and a tolerance factor of 0.01~\cite{Feroz:2007kg}, which ensure that the sampling is accurate enough to have a parameter estimation similar to Markov-Chains Monte Carlo sampling methods.

\section{Direct Detection rates and interaction scenarios $\mathcal{M}_i$}\label{sec:dd}

The differential spectrum for a nuclear recoil due to scattering of a WIMP, in units of  events per time per detector mass per energy, 
has the form
\begin{equation}
\label{eq:diffrate}
\frac{{\rm d}R}{{\rm d}E} = \frac{\rho_{\odot}}{m_{\rm DM} }  \int_{v'>v'_{\rm min}} {\rm d}^3v'  \, \frac{{\rm d}\sigma}{{\rm d}E} \, v'  \, f (\vec{v'}(t))\,,
\end{equation}
where $E$ is the energy transferred during the collision,  $\rho_{\odot} \equiv \rho_{\rm DM}(R_{\odot})$ the WIMP density in the solar neighbourhood,
$m_{\rm DM}$ the dark matter mass and
${\rm d}\sigma/{\rm d}E$ the differential cross section for the scattering. $ f(\vec{v'}(t))$ is 
the WIMP velocity distribution in the Earth's rest frame normalised such that $\int {\rm d}^3 v' f(\vec{v'}(t))=1$, which we describe in section~\ref{sec:veldistr}.

The total number of recoils expected  in a detector in a given observed energy range $[{\cal E}_1,{\cal E}_2]$ is obtained by integrating equation~(\ref{eq:diffrate}) over energy
\begin{equation}
\label{eq:totrate}
S(t)=M_{\rm det} T \int_{{\cal E}_1/q}^{{\cal E}_2/q}  {\rm d}E\  \epsilon(q E)\  \frac{\rm dR}{{\rm d}E} \,,
\end{equation}
where $M_{\rm det}\, T$ denotes the detector total mass times the exposure time. We have folded into the integral an energy-dependent function $\epsilon(q E)$ describing the efficiency of the detector. The quenching factor $q$, defined via ${\cal E}=q E$, denotes the fraction of recoil energy that is ultimately observed in
a specific detection channel and is a detector-dependent quantity. To distinguish ${\cal E}$ from the 
actual nuclear recoil energy $E$, the former is usually given in units of keVee (electron equivalent keV), while the latter in keVnr (nuclear recoil keV) or simply keV.

In our analysis, we consider spin-independent (SI) scattering off nuclei, encoded in the differential cross-section in the following way: 
\begin{equation}
\label{eq:pppart}
\frac{\rm d\sigma}{{\rm d}E} = \frac{M_{\cal N} \sigma^{\rm SI}_n}{2 \mu^2_n {v}'^2}\ \frac{\Big(f_p Z + (A-Z) f_n\Big)^2}{f_n^2} {\cal F}^2( E) \,  ,
\end{equation}
where $\mu_n=m_{\rm DM} m_n/(m_{\rm DM}+m_n)$ is the WIMP-nucleon reduced mass, $\sigma^{\rm SI}_n$ the spin-independent zero-momentum WIMP-nucleon cross-section, $Z$ ($A$) the atomic (mass) number of the target nucleus used, and $f_p$ ($f_n$) is the WIMP effective coherent coupling to the proton (neutron). The nuclear form factor ${\cal F}(E)$ characterises the loss of coherence for nonzero momentum transfer: a fair approximation for all nuclei is the Helm form factor~\cite{Helm:1956zz,*Lewin:1995rx}. We consider three hypotheses for the type of interaction, further treated as nested models $\mathcal{M}_i$, as follows.

\begin{enumerate} 
\item Elastic scattering ($\mathcal{M}_0$)

This is the standard interaction common to many WIMP models. In practice it consists in the following assumptions. The integration in equation~\ref{eq:diffrate} is performed over all incident particles capable of depositing a recoil energy of $E$, which implies a lower integration limit of $v'_{\rm min} = \sqrt{M_{\cal N} E/2 \mu}$, where $M_{\cal N}$ is the mass of the target nucleus, and $\mu=m_{\rm DM} M_{\cal N}/(m_{\rm DM}+M_{\cal N})$ is the WIMP-nucleus reduced mass. In equation~\ref{eq:pppart} we set $f_n=f_p$, that is same coupling to neutron and proton and consequently the interaction scales as usual as $A^2$. This corresponds to scalar interaction, {\it e.g.} DM scattering off nucleons exchanging a Higgs boson. There are two theoretical parameters for the WIMP interaction: $m_{\rm DM}$ and $\sigma_n^{\rm SI}$. For model comparison this is the simplest model, called hereafter $\mathcal{M}_0$. 

\item Inelastic scattering ($ \mathcal{M}_1$)~\cite{TuckerSmith:2001hy}

A WIMP $\chi$ may scatter off nuclei only by making a transition into an heavier state: $\chi \mathcal{N} \to \chi^{\ast} \mathcal{N}$.  The two DM mass eigenstates have a mass splitting proportional to $\Delta m \equiv \delta$, which is of the order of $\mathcal{O}({\rm keV})$ in order to the scatter to occur. Only particles in the very high tale of the velocity distribution will have enough energy to produce a recoil in the detector, that translates into a modified minimal scattering velocity:
\begin{equation}
v'_{\rm min} = \sqrt{\frac{1}{2 M_{\mathcal N} E_R}} \Big(\frac{M_{\mathcal N} E_R}{\mu_n}+\delta\Big)  \,.
\label{eq:vmininel}
\end{equation}
Heavy nuclei will be particularly sensitive to this interaction, therefore for this scenario we do not consider data on Si, F and Cl. There are 3 free parameters: same as in $\mathcal{M}_0$ plus the mass splitting $\delta$, which we vary with a flat prior between 0 (elastic limit) to 200 keV. This model is denoted $\mathcal{M}_1$ in Bayesian comparison.

\item Isospin violating scattering ($\mathcal{M}_2$)~\cite{Feng:2011vu} 

This model relies on the hypothesis that the WIMP interaction with the neutron and the proton might be of different strength, namely $f_n \neq f_p$ in equation~\ref{eq:pppart}. The minimal velocity is defined as for the elastic interaction. The SI cross-section is the mean between the one on neutron and the one on proton:
\begin{equation}
\sigma^{\rm SI} = \frac{\sigma^{\rm SI}_n+\sigma^{\rm SI}_p}{2}\,.
\end{equation}
Different nuclei isotopes, each with abundance $r_i$ in the detector, are taken into account replacing the $A^2$ factor with an effective one:
\begin{equation}
A^2_{\rm eff} = \sum_{i = isotopes} 2 r_i \left[Z f_p + (A_i - Z) f_n\right]^2\, ,
\end{equation}
following~\cite{Schwetz:2011xm}.
There are 3 free parameters: the two as in $\mathcal{M}_0$ plus $f_n/f_p$. We let free to vary this ratio from -2 (an asymptotic limit at which all nuclei behave the same) to 1 (elastic scattering limit) with a flat prior, not to favour any value in particular. This model will be referred as $\mathcal{M}_2$.

\end{enumerate}

The parameters describing the WIMP interaction in each model are resumed together with their prior range in table~\ref{tab:prior2}. The choice for flat/log priors we follow here has been discussed in~\cite{Arina:2011si}.

\begin{table}[t!]
\caption{\texttt{MultiNest} parameters and priors for the WIMP parameter space in the three models of SI interaction considered in this work. All priors are uniform over the indicated range.\label{tab:prior2}}
\begin{center}
\begin{ruledtabular}
\begin{tabular}{lll}
Model & Parameter & Prior \\
\hline
All & $\log(m_{\rm DM}/{\rm GeV})$  & $0  \to 3$\\
All & $\log(\sigma_n^{\rm SI}/{\rm cm}^2)$ & $-46 \to -36$\\ 
Inelastic ($\mathcal{M}_1$)& $\delta/({\rm keV})$  &  $0 \to 200$ \\
Isospin violating ($\mathcal{M}_2$) & $f_n/f_p$ & $-2 \to 1$  \\
\end{tabular}
\end{ruledtabular}
\end{center}
\end{table}

\section{Likelihood definition}\label{sec:expdlkh}

In this section we shortly define the likelihood function for CRESST, KIMS and bubble chamber experiments. We review the likelihood for Xe100, in light of the recent data~\cite{:2012nq} as well. For DAMA and CDMS on Silicon we use the set up defined in~\cite{Arina:2011si}, while for CoGeNT we use the publicly available data, see~\cite{Arina:2011zh}. We do not consider CDMS data on Ge, that have been discussed extensively in~\cite{Arina:2011si}, since they are less constraining than other exclusion bounds considered in this analysis. We do not consider the low energy analyses by XENON10~\cite{Angle:2011th} and CDMS~\cite{Ahmed:2010wy,*Akerib:2010pv}, as well as the modulated analysis by CDMS~\cite{Ahmed:2012vq} because of the lack of a reliable parametrization of the background making difficult the construction of a meaningful likelihood function for our Bayesian analysis. 

We resume all the experiments we consider with their nuisance parameters, due to systematics, and their prior range in table~\ref{tab:prior1}. The details about likelihood construction are presented in appendix~\ref{sec:appa}. At the end of the section we briefly recall how nuisance parameters coming from astrophysics are implemented.

\subsection{Experimental likelihoods}

\paragraph*{XENON100} 
The likelihood $\ln \mathcal{L}_{\rm Xe100}$ is defined in~\cite{Arina:2011si}, implemented however with the latest data. The last scientific run has observed 2 events ($N_{\rm obs} = 2$). Actually it is precisely $N_{\rm obs}$ that will be tested under $\mathscr{L}$ and $\mathscr{R}$-tests. We will compute the joint evidence for Xe100 and the combined set $\{{\rm DAMA, CRESST ,CoGeNT}\}$, as in equation~\ref{eq:jointpdf}. For this purpose we scan over a finite number of realizations under the variable $N_{\rm events}$:
\begin{equation}
N_{\rm events}: {\rm 0,\,  10,\,  ...,\,  60 \, (100)\,  in \, intervals\,  of \, 10}
\end{equation}
plus the evaluation of the joint evidence at $N_{\rm events} = N_{\rm obs}$. We choose the maximum numbers of events that can be seen by Xe100 in 225 live day of run to be 60, which is reasonable compared to the forecasts in~\cite{Pato:2010zk,Strege:2012kv}. We then interpolate between data points with a spline to get the joint evidence as a function in data space. 

\paragraph*{CRESST}  The likelihood is constructed on the total number of events seen in each detector module and on the background modelling given in section 4 of~\cite{Angloher:2011uu}. The yield information is not included in the analysis. The backgrounds constitute the nuisance parameters, over which we marginalise.
 
\paragraph*{Bubble chamber experiments} We consider PICASSO~\cite{Archambault:2012pm}  and SIMPLE, phase II~\cite{Felizardo:2011uw}. These detectors capture phase transitions produced by the energy deposition of a charged particle traversing the liquid, if the generated heat spike occurs within a certain critical length and exceeds a certain critical energy. The event is accompanied by an acoustic signal. Therefore the detectors perform as threshold devices, controlled by setting the temperature $T$ and/or the pressure. The relation between the energy threshold $E_{\rm th}(T)$ and the temperature is obtained at a fixed pressure during the calibration process. The observed rate per day per kg of material is then defined as:
\begin{eqnarray}\label{eq:rsupheat}
S
& = &  \int_0^{E_{\rm max}} {\rm d}E \, P(E, E_{\rm th}(T)) \,  \frac{{\rm d} R}{{\rm d} E}\,,
\end{eqnarray}
where $E_{\rm max}$ is the maximum energy released by a WIMP with a certain escape velocity $v_{\rm esc}$ and $P(E, E_{\rm th}(T))$ describes the effect of a finite resolution at threshold, approximated by:
\begin{equation}
P(E, E_{\rm th}(T)) = 1 - \exp \left[ a(T) \left(1 -\frac{E}{E_{\rm th}(T)}\right)\right]\,.
\end{equation}
The parameter $a(T)$ defines the steepness of the energy threshold, and is a nuisance parameter for both experiments. The details on the remaining of the likelihood are given in the appendix~\ref{sec:appa} for each collaboration separately. 

\begin{table}[t!]
\caption{\texttt{MultiNest} parameters and priors for experimental systematics (nuisance parameters).  All priors
are uniform over the indicated range.\label{tab:prior1}}
\begin{center}
\begin{ruledtabular}
\begin{tabular}{lll}
Experiment&  Parameter & Prior \\
\hline
DAMA & $q_{\rm Na}$ &  $0.2 \to 0.4$\\
DAMA & $q_{\rm I}$ &  $0.06 \to  0.1$\\
CoGeNT & $C$ &  $0 \to 10$~cpd/kg/keVee \\
CoGeNT & ${\cal E}_0$ &  $0 \to 30$~keVee\\
CoGeNT & $G_n$ & $0 \to 10$~cpd/kg/keVee\\
CRESST & $N_\alpha$ &  $5 \to 17$ counts\\
CRESST & $C_{\rm Pb}$ & $1 \to 7$  counts/keV\\
CRESST & $N_n$ &  $3.3 \to 34$ counts\\
Xe100 & $m$ & $-0.01 \to 0.18$\\
PICASSO & $a(T)$ & $1 \to 11$  \\
SIMPLE & $a(T)$ & $1 \to 11$ \\
KIMS & $q_I$ &  $0.06 \to 0.1$ \\
KIMS & $q_{Cs}$ &  $0.06 \to 0.1$\\
KIMS & $B_{\alpha}$ & $0 \to 0.4$\\
\end{tabular}
\end{ruledtabular}
\end{center}
\end{table}

\paragraph*{KIMS} This experiment~\cite{Kim:2012rz} has a binned Gaussian likelihood for describing the counts/keV/kg/day seen in the detectors, which are compatible with the no detection hypothesis. In addition it has three nuisance parameters from $\alpha$ background and quenching factors.

\subsection{Astrophysical uncertainties}\label{sec:veldistr}

\begin{table}[t!]
\caption{Astrophysical constraints on the DM halo profile and the WIMP velocity distribution.\label{tab:prior3}}
\begin{center}
\begin{ruledtabular}
\begin{tabular}{ll}
Observable & Constraint  \\
\hline
Local standard of rest&  $v_0^{\rm obs} = 230 \pm 24.4 \ {\rm km \ s}^{-1}$~\cite{Bovy:2012ba,*Reid:2009nj,*Gillessen:2008qv}\\
Escape velocity &   $v_{\rm esc}^{\rm obs}= 544   \pm 39 \  {\rm km \ s}^{-1}$~\cite{Smith:2006ym,*Dehnen:1997cq} \\
Local DM density & $\rho_{\odot}^{\rm obs} = 0.4 \pm 0.2 \  {\rm GeV \ cm}^{-3}$~\cite{Weber:2009pt,*Salucci:2010qr,*Bovy:2012tw} \\
Virial mass &  $M_{\rm vir}^{\rm obs} = 2.7  \pm 0.3 \times 10^{12} M_{\odot}$~\cite{Dehnen:2006cm,*Sakamoto:2002zr} \\
\end{tabular}
\end{ruledtabular}
\end{center}
\end{table}

As for the WIMP velocity distribution entering in the rate equation~\ref{eq:diffrate}, we consider two alternatives. For details we refer to~\cite{Arina:2011si,Arina:2011xu}.

\begin{enumerate}
\item The standard halo model (SMH) 

It is commonly used in direct detection prediction for extracting experimental bounds and consists in a Maxwellian distribution with fixed astrophysical parameters $v_0$, $v_{\rm esc}$ and $\rho_\odot$. We choose to fix the parameters at their mean value, as given in table~\ref{tab:prior3}. It allows to clearly visualise the sensitivity of exclusion bounds/detection regions on experimental systematics.

\item DM density profile (NFW)

We construct self consistent halo distributions starting from a motivated DM density profile, the NFW halo distribution~\cite{Navarro:1996gj}, as shown in~\cite{Arina:2011si}. The DM density profile is constructed from the virial mass $M_{\rm vir}$ and the concentration parameter $c_{\rm vir}$. Then by means of the Eddington formula we extract the corresponding velocity distribution. We marginalise over the nuisance parameters $M_{\rm vir}, c_{\rm vir}, v_{0}, v_{\rm esc}$ and $\rho_\odot$. The astrophysical likelihood is given by
\begin{eqnarray} 
\label{eq:lkhastro}
& \ln{\cal L}_{\rm Astro}  =  \! -  \frac{(v_0 - \bar{v}^{\rm obs}_0)^2}{2 \sigma^2_{v_0}} \! -  \! \frac{(v_{\rm esc} - \bar{v}^{\rm obs}_{\rm esc})^2}{2 \sigma^2_{v_{\rm esc}}} \!\nonumber\\
& -  \! \frac{(\rho_\odot - \bar{\rho}^{\rm obs}_\odot)^2}{2 \sigma^2_{\rho_\odot}}\!-  \! \frac{(M_{\rm vir} - \bar{M}^{\rm obs}_{\rm vir})^2}{2 \sigma^2_{M_{\rm vir}}} - C_{\rm norm}\,,
\end{eqnarray}
with gaussian prior centered on the experimental measured values quoted in table~\ref{tab:prior3}. The normalization factor $C_{\rm norm} = \ln (2 \pi \sigma^2_{v_0}) + \ln (2 \pi \sigma^2_{\rho_\odot}) + \ln (2 \pi \sigma^2_{v_{\rm esc}}) + \ln (2 \pi \sigma^2_{M_{\rm vir}})$ is fundamental for computing the evidence.

Other DM density profiles give similar results on the \{$m_{\rm DM},\sigma_n^{\rm SI}$\}-plane, namely the exact shape of the DM halo density profile, at least within the class of spherically symmetric, smooth profiles, does not yet play a role in direct DM searches, as shown in~\cite{Arina:2011si}. Even if it does not capture completely the distribution in the galaxies of DM particles~\cite{Donato:2009ab}, it is a fair approximation to consider a NFW density profile.
\end{enumerate}

\section{State of the art}\label{sec:soa}

The present situation of direct detection experiments is shortly illustrated for the three spin-independent interaction models we consider in this work. For more details we refer to appendix~\ref{sec:appb}.

\paragraph*{Elastic SI scattering (model $\mathcal{M}_0$)}
\begin{figure*}[t]
\includegraphics[width=0.49\textwidth,trim=20mm 65mm 20mm 65mm, clip]{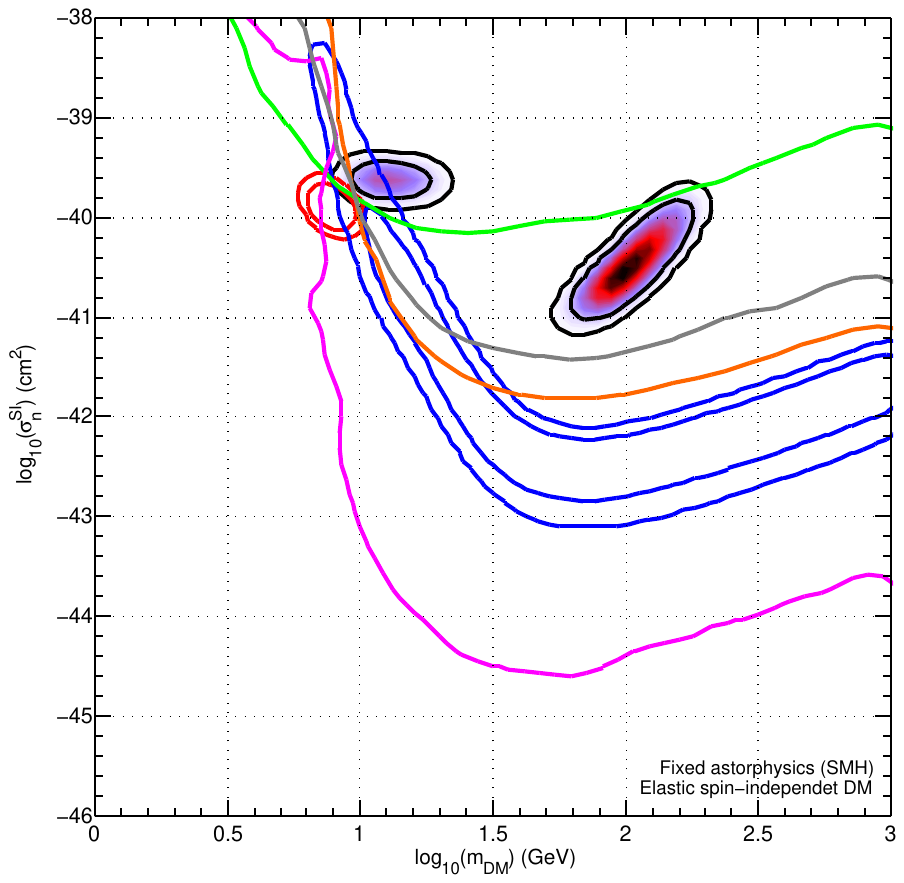}
\includegraphics[width=0.49\textwidth,trim=20mm 65mm 20mm 65mm, clip]{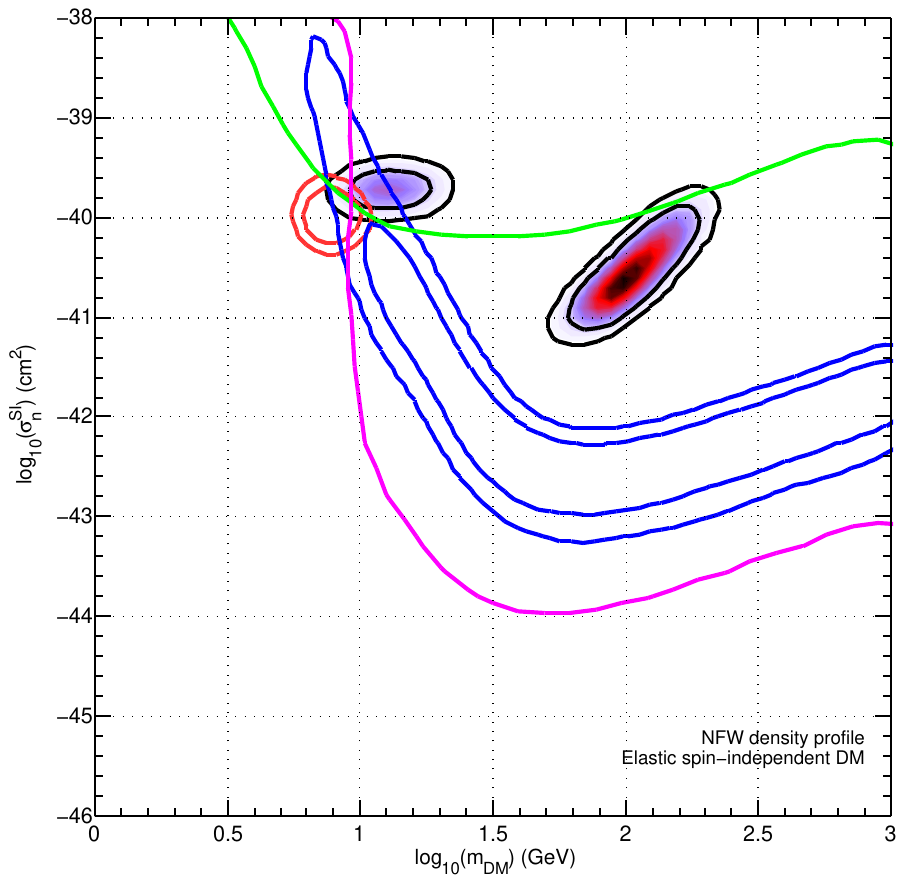}
\caption{ Elastic spin-independent (SI) scattering. {\it Left}: 2D credible regions for the individual experimental bounds and regions assuming fixed astrophysical parameters (SMH), combined in a single plot. For DAMA (shaded), CoGeNT (red)  and CRESST (blue) we show the 90\% and 99\% contours. The orange line represents the $90_S$\% bound for CDMSSi, the magenta curve is for Xe100, the light gray curve stands for SIMPLE, the light green is for PICASSO. The experimental nuisance parameters are marginalised over. {\it Right:} Same as left with the marginalisation over the astrophysical uncertainties using a NFW density profile for the dark matter.  Only Xe100 (magenta solid) and PICASSO (light green solid) are shown, because they are the most constraining ones for the considered scenario.\label{fig:All_SMH}}
\end{figure*}

\begin{table}[t]
\caption{1D posterior pdf modes and $90\%$ credible intervals for the circular velocity $v_0$,
escape velocity $v_{\rm esc}$, and the local DM density $\rho_{\odot}$ for NFW density profile considered in this work and for the elastic ($\mathcal{M}_0$), inelastic ($\mathcal{M}_1$) and isospin violating ($\mathcal{M}_2$) scenarios. \label{tab:astro}}
\begin{center}
\begin{ruledtabular}
\begin{tabular}{ l | lll }
& $v_0$ (${\rm  km\,s}^{-1}$) & $v_{\rm esc}$ (${\rm km\, s}^{-1}$) & $\rho_{\odot}$  ($ {\rm GeV\,cm}^{-3}$) \\ 
\hline
 $\mathcal{M}_0$ & & & \\
DAMA & $220^{+40}_{-20}$ & $558_{-16}^{+19}$ & $0.37_{-0.09}^{+0.15}$ \\
CoGeNT & $219_{-18}^{+38}$ & $559 \pm 17$ & $0.37_{-0.08}^{+0.20}$ \\
CRESST & $221_{-18}^{+40}$ & $558_{-16}^{+19}$ & $0.38_{-0.10}^{+0.15}$ \\
PICASSO & $221_{-21}^{+40}$ & $558_{-18}^{+20}$ & $0.38_{-0.10}^{+0.15}$ \\
Xe100 &$221_{-24}^{+38}$ &$558_{-16}^{+19}$ & $0.40_{-0.12}^{+0.13}$ \\
\hline
$\mathcal{M}_1$ & & & \\
DAMA & $221^{+34}_{-19}$ & $558_{-15}^{+19}$ & $0.38_{-0.08}^{+0.15}$ \\
CoGeNT & $225_{-19}^{+42}$ & $558_{-16}^{+22}$ & $0.40_{-0.08}^{+0.16}$ \\
CRESST & $222_{-19}^{+41}$ & $558_{-17}^{+20}$ & $038._{-0.10}^{+0.15}$ \\
KIMS & $220_{-21}^{+41}$ & $558_{-18}^{+22}$ & $0.38_{-0.10}^{+0.16}$ \\
Xe100 &$223_{-23}^{+37}$ &$558_{-17}^{+20}$ & $0.39_{-0.11}^{+0.14}$ \\
\hline
$\mathcal{M}_2$ & & & \\
DAMA & $220^{+38}_{-18}$ & $558_{-15}^{+19}$ & $0.38_{-0.09}^{+0.14}$ \\
CoGeNT & $219_{-21}^{+38}$ & $557_{-16}^{+19} $ & $0.37_{-0.09}^{+0.16}$ \\
CRESST & $222_{-23}^{+39}$ & $558_{-17}^{+20} $ & $0.38_{-0.09}^{+0.15}$ \\
PICASSO & $221_{-21}^{+40}$ & $558_{-18}^{+20}$ & $0.38_{-0.10}^{+0.15}$ \\
Xe100 &$222_{-22}^{+37}$ &$558_{-17}^{+21} $ & $0.39_{-0.11}^{+0.14}$ \\
\end{tabular}
\end{ruledtabular}
\end{center}
\end{table}

\begin{figure*}[t]
\includegraphics[width=0.49\textwidth,trim=20mm 65mm 20mm 65mm, clip]{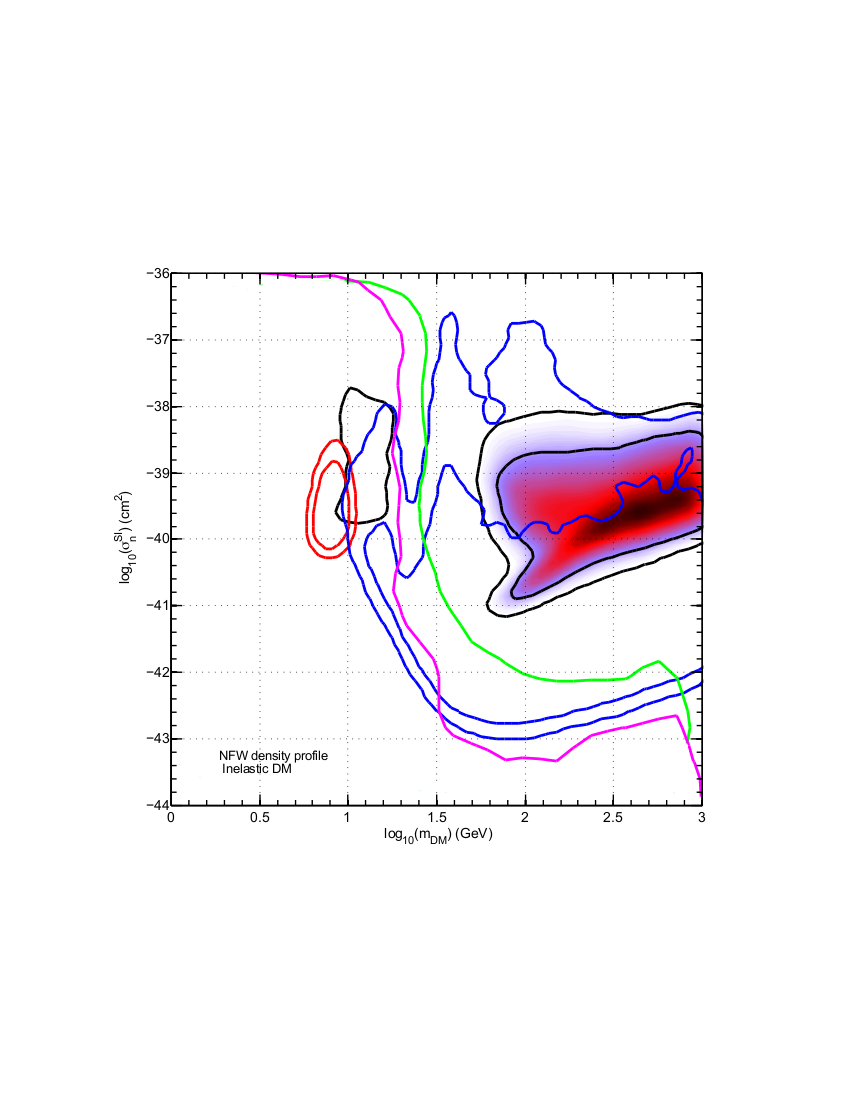}
\includegraphics[width=0.49\textwidth,trim=20mm 65mm 20mm 65mm, clip]{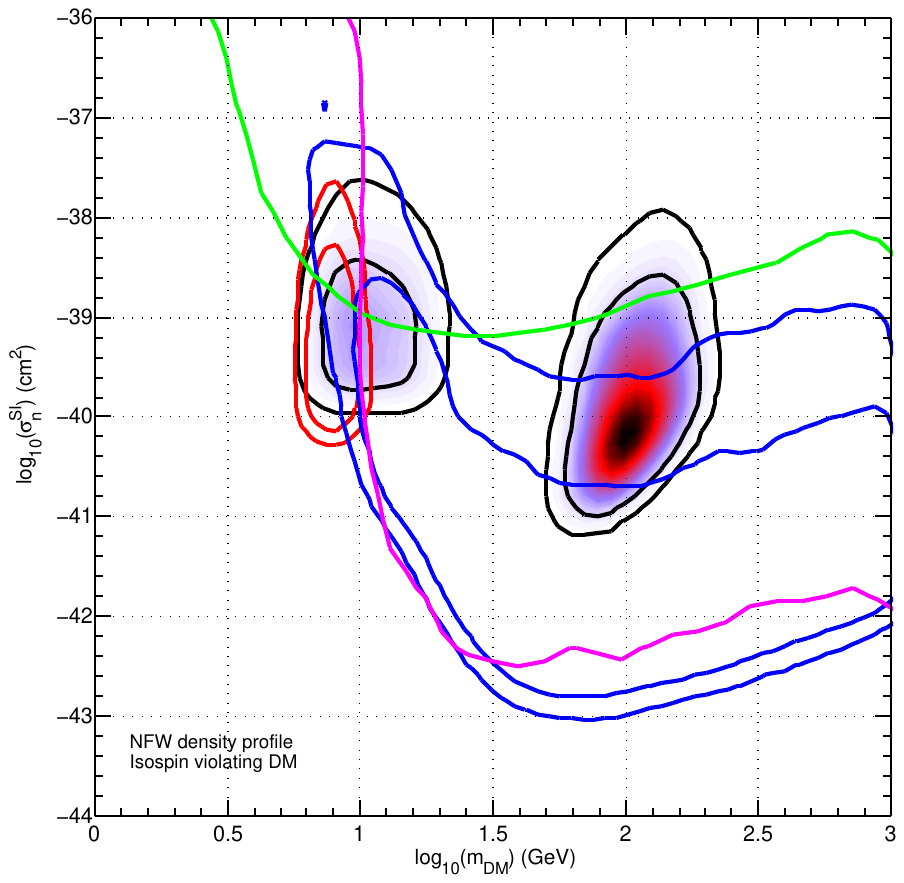}
\caption{{\it Left:} Same as figure~\ref{fig:All_SMH} for inelastic SI interaction. The green dark (magenta) curve is the KIMS (Xe100) $90_S\%$ exclusion bound. The mass splitting direction $\delta$ has been marginalised over. {\it Right:} Same as figure~\ref{fig:All_SMH} for isospin violating interaction. The light green (magenta) line denotes the PICASSO (Xe100) exclusion limit at $90_S\%$ confidence level. The isospin violating parameter $f_n/f_p$ has been marginalised over. In both panels the astrophysical uncertainties are marginalised over, considering a NFW density profile for the dark matter, as well as all the experimental systematics. Only the most constraining exclusion limits are shown. Labelling for the closed regions is as in figure~\ref{fig:All_SMH}. \label{fig:All_NFW}}
\end{figure*}

The 2D marginal posterior pdf in the $\{m_{\rm DM},\sigma^{SI}_n\}$-plane for all the individual experiments is combined in a single plot in figure~\ref{fig:All_SMH}. We first consider the left panel, where the astrophysical quantities are fixed at their mean value and only the effects of marginalising over systematics appear. One can easily recognise the DAMA credible region (shaded), the CoGeNT one (red non filled) and the CRESST region (blue non filled) with contours at $90\%$ and $99\%$. All exclusion bounds are at $90_S\%$ confidence level. By means of the `chi by eye', it is apparent that DAMA and CRESST are disfavoured at $90_S\%$ by Xe100, while CoGeNT is still partially compatible. On the same foot the PICASSO upper limit challenges DAMA, which is incompatible at $90_S\%$, while being compatible with CoGeNT. All other exclusion limits (as labelled in the caption) are less relevant for the elastic spin-independent scenario. None of the nuisance parameters show an interesting behavior.

The right panel of figure~\ref{fig:All_SMH} displays the case of a velocity distribution constructed starting from a NFW halo profile for the dark matter with marginalisation over the astrophysical parameters, in addition to the systematics. Firstly, we note that  allowing for uncertainties in the astrophysics significantly expands the closed regions of DAMA, CoGeNT and CRESST, while the exclusion limits  tend to shift a little to the right. This increases the compatibility: DAMA, CoGeNT and CRESST credible regions overlap now within their 90\% contours and are partially compatible with both Xe100 and PICASSO at $90_S\%$. Secondly we note that direct DM searches are not at the moment contributing towards constraining the astrophysics of the problem. Indeed for a given DM halo profile the preferred values for $v_0$, $v_{\rm esc}$ and $\rho_\odot$ and their associated uncertainties are virtually independent of the additional constraints from the DM experiments. As a consequence the experimental systematics follow the same trend as for SMH case. For a given DM density profile, the preferred value for the astrophysical parameter is very similar in all the three spin-independent scenarios, as confirmed by table~\ref{tab:astro}: an insight on the astrophysical properties of the DM by means of particle physics (and vice versa) appears beyond the current potential of direct searches. 

In the light of the above considerations, we present the other interaction models marginalised over the astrophysics.

\paragraph*{Inelastic SI scattering (model $\mathcal{M}_1$)} 

The summary in a single plot of all individual experimental outcomes is given in figure~\ref{fig:All_NFW} (left panel) as a function of the dark matter mass and scattering cross-section. Same labelling as for elastic SI case for detection regions; in this case the most constraining experiments are Xe100 (magenta) and KIMS (green), the only ones shown in the plot. The usual Iodine region for DAMA is excluded at $90_S\%$ by both experiments, however  there is room for a consistent explanation at low WIMP mass at $90_S\%$ confidence level. This is again a `chi by eye' consideration, and we will show that Bayesian model comparison may come out with different results, because of the Occams' razor principle.
The exclusion bounds and detection regions are affected by a volume effect not only due to astrophysical marginalisation but also due to marginalisation over the mass splitting parameter $\delta$. In appendix~\ref{sec:appb} the experimental dependence on it is detailed.

\paragraph*{Isospin violating SI interaction (model $\mathcal{M}_2$)}

The right panel of figure~\ref{fig:All_SMH} illustrates the state of the art for isospin violating SI scattering (contours/lines labelling in the caption). All the three detection regions overlap for $\sigma_n^{SI} \sim 10^{-39} {\rm cm^2}$ and a DM mass of 10 GeV: the data are compatible at 90\% confidence level. The closed contours again are enlarged by volume effects due to marginalisation over the isospin violating parameter $f_n/f_p$. Moving on the exclusion bounds we see immediately that Xe100 is the most constraining experiment for DM masses above 15 GeV while below that value it does find common ground for DAMA, CoGeNT and CRESST. This is by virtue of the isospin violating interaction, which depletes the interaction on Xe whit respect to Na or partially Ge in a certain range of $f_n/f_p \sim -0.7$. The low mass regions of DAMA, CRESST and CoGeNT are compatible with the $90_S\%$ upper bound of PICASSO as well. By means of the `chi by eye', we could conclude, as in the case of inelastic SI scattering, that this particle physics scenario accomplishes a better agreement between individual detection regions among themselves and with the exclusion bounds than the elastic SI scenario. We might want to confront these statements with the outcomes of Bayesian model selection.

In conclusion at present Xe100 is the exclusion bound that really challenges the detection regions in all the SI scenarios we have considered. In the next section  we assess rigorously at which statistical significance they are (in)consistent within each other.

\section{Results and discussion}\label{sec:res}

Here we describe the outcomes for Bayesian consistency tests between Xe100 and the detection regions, section~\ref{sec:tests}. We will find that in all scenarios but inelastic SI model the inconsistency is at the level of 2$\sigma$. It is therefore not interesting neither meaningful to attempt a global fit: we limit at the detection regions the investigation on how direct detection data can constrain particle physics models, in section~\ref{sec:modcomp}.

\subsection{Consistency tests}\label{sec:tests}

\begin{figure*}[t]
\includegraphics[width=0.49\textwidth,trim=25mm 60mm 30mm 60mm, clip]{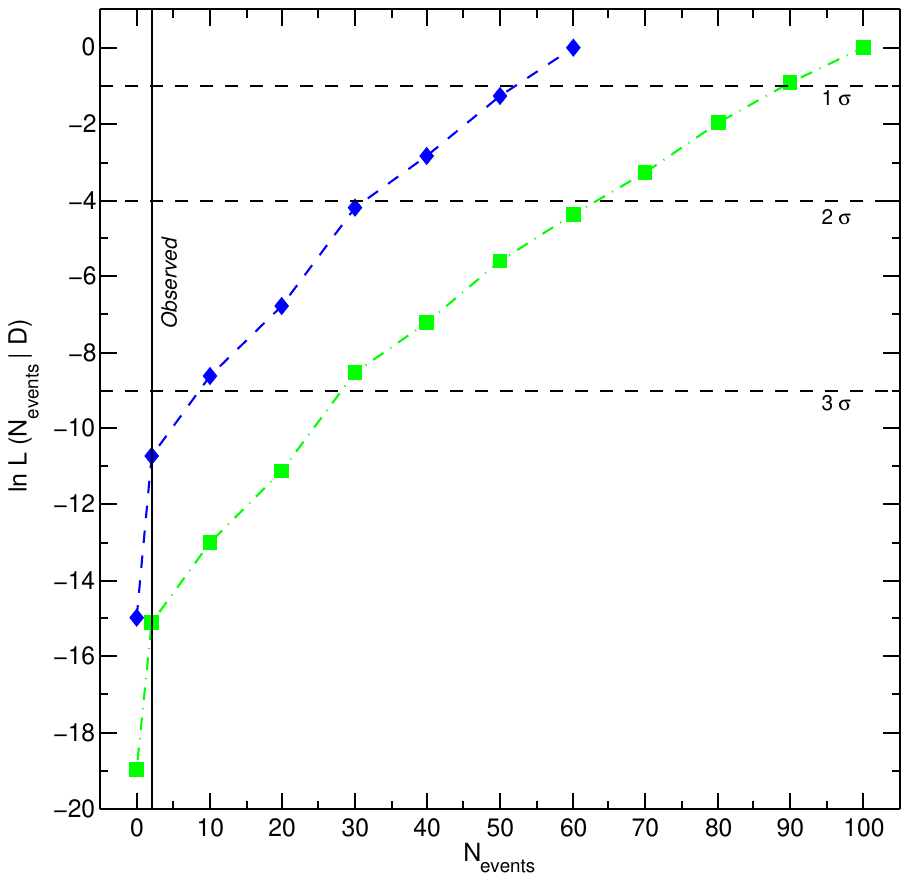}
\includegraphics[width=0.49\textwidth,trim=25mm 60mm 30mm 60mm, clip]{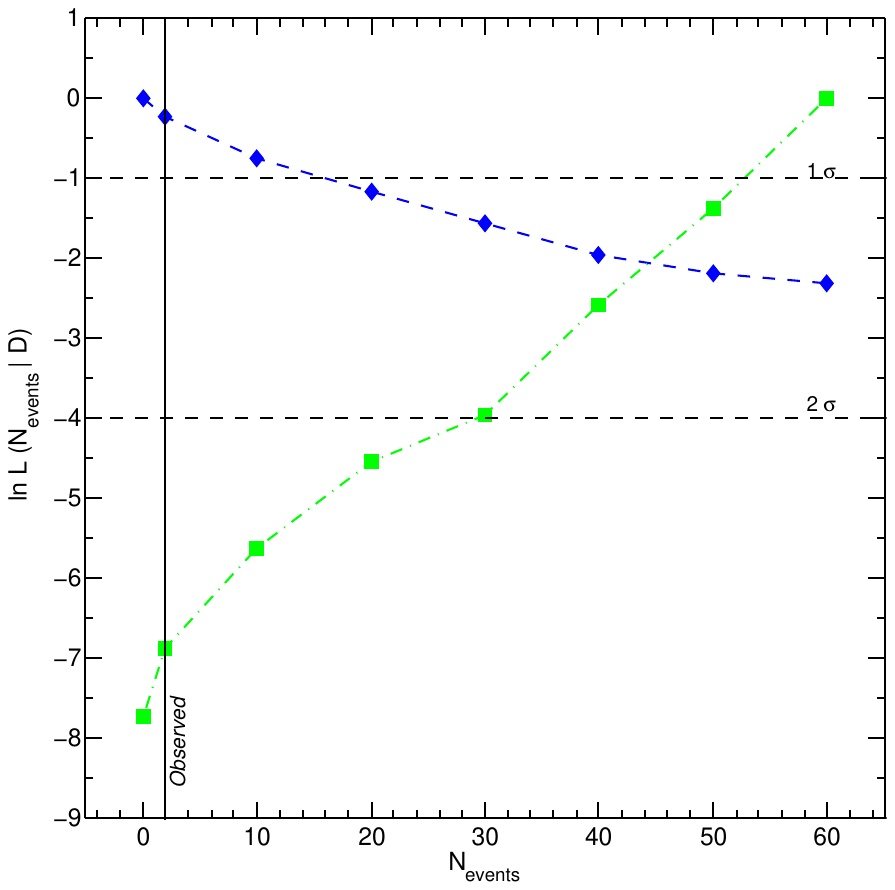}
\caption{{\it Left}: Predictive data distribution ($\mathscr{L}-$test) for the number of events $N_{\rm events}$ in Xe100 detector, as defined in equation~\ref{eq:ltest}. The curve represents the conditional evidence of Xe100 and the combined set ($D=\{{\rm DAMA, CoGeNT, CRESST}\}$) at a given data point, divided by the maximum of the probability, for elastic SI interaction. The blue dashed curve is for a maximum of 60 events, while the green dot-dashed line stands for $N_{\rm max}=100$. The vertical line gives the actual measured value $N_{\rm obs}=2$. The data points denote the location at which the predictive probability has been computed and the lines are spline interpolation between those points. The horizontal dashed line represents the $1,2$ and $3\sigma$ significance. {\it Right: } Same as left for inelastic ($\mathcal{M}_1$) and isospin violating ($\mathcal{M}_2$) scenarios, blue and green data points/line respectively. In these scenarios we assumed $N_{\rm max}=60$. All astrophysical uncertainties and experimental systematics have been marginalised over, as well as all the model parameters. \label{fig:TestEDM}}
\end{figure*}

Regarding the assessment of compatibility between the data sets $\mathscr{D}\equiv N_{\rm events}$ and $D = \{{\rm DAMA, CoGeNT, CRESST}\}$, we present our predictions in data space and not anymore in the model parameter space, because of the definition of equation~\ref{eq:jointpdf}. 

We first discuss the $\mathscr{L}$-test. We have considered different possible outcomes for the observed number of events in the Xe100 detector, with fixed instrumental noise as reported by the collaboration, which is a reasonable assumption. We have evaluated the conditional evidence $p(\mathscr{D}|D)$ and computed the predictive probability on a grid of values for $N_{\rm events}$. The relevant quantity $\mathscr{\ln L}(N_{\rm events}|D)$ is plotted in figure~\ref{fig:TestEDM} as a function of the possible outcome of the experimental observation, with the actual observed value denoted by a solid black vertical line. The elastic SI scattering is given in the left panel. Consider first the blue line/diamonds: the predictive probability grows fast increasing the number of events seen in the detector. This indicates that actually the compatibility of this experiment with $D$ increases augmenting the number of events seen in Xe100. In other words a number of events larger than 2 should have been observed for $\mathscr{D}$ and $D$ to be consistent. We see in addition that the maximum of the probability depends on the maximum number of events we assume have been seen. Considering $N_{\rm max}=60$ the discrepancy between the data sets $D$ and $\mathscr{D}$ is larger than 3$\sigma$. Augmenting the number of `observed' events in the detector (green line and square, with $N_{\rm events}=100$) would lead to a even larger discrepancy. In the right panel, the predictive probability for the inelastic SI scattering scenario (blue/diamonds) has the opposite behavior than $\mathcal{M}_0$: the finest agreement between Xe100 and the combined fit is found for 0 observed events. This actually is supported by the parameter inference (discussed below) because the combined fit $D$ favours the low DM mass, while Xe100 inelastic is unable to exclude such region. Therefore augmenting the observed number of events leads to an increasing inconsistency. We conclude that for inelastic interaction Xe100 is compatible  within 1$\sigma$ with DAMA, CoGeNT and CRESST and this significance is robust against the assumed value of $N_{\rm max}$. The isospin violating SI scenario (green/squares) follows closely the behavior of elastic scattering, although the discrepancy in that case is marginal, at the level of 2$\sigma$, for $N_{\rm max}=60$. 

Note that this probability distribution does not make advantage of the spectral information of the $N_{\rm events}$ in the likelihood ({\it e.g.} for a light WIMP the events should be concentrated in the low energy part of the detection range) and keeps growing by increasing the number of observed events. It can be taken therefore as a conservative assessment of significance, that may be reduced by allowing this extra information. The (in)consistency between Xe100 and $D$ in the isospin violating scenario may be lowered to $1\sigma$ level assuming at most 20 events in the detector. For the same number of events and elastic SI picture, the experimental data sets are still incompatible but with a statistical significance of only $2 \sigma$.

\begin{table}[t]
\caption{Results for the $\mathscr{R}-$test, providing the relative odds between the consistency hypothesis $\mathcal{H}_0$ (Xe100 and $D=\{{\rm DAMA, CoGeNT, CRESST}\}$ consistent with each other) and the incongruous belief of $\mathcal{H}_1$ (Xe100 and $D$ inconsistent). $\mathcal{H}_0 $ is favoured for $\ln \mathscr{R} > 0$, while the data sets are in tension with each other for $\ln\mathscr{R}<0$. We give the test results in the three particle physics scenarios under investigation, as labelled. The statistical interpretation is in accordance with Jeffreys' scale, given in table~\ref{tab:jef} and the definition of the $\mathscr{R}-$test is given in equation~\ref{eq:rtest}. All astrophysical uncertainties and experimental systematics have been marginalised over, as well as the model parameters. \label{tab:Rtest}}
\begin{center}
\begin{ruledtabular}
\begin{tabular}{c|c|c}
Model & $\ln \mathscr{R}(N_{\rm obs}=2)$ & Interpretation \\
\hline
$\mathcal{M}_0$ & $-0.32 \pm 0.07$  & Inconclusive evidence against $\mathcal{H}_0$\\
$\mathcal{M}_1$ & $-0.53\pm 0.07$ & Inconclusive evidence against $\mathcal{H}_0$  \\
$\mathcal{M}_2$ &$-0.22\pm 0.07$ & Inconclusive evidence against $\mathcal{H}_0$ \\
\end{tabular}
\end{ruledtabular}
\end{center}
\end{table}

The $\mathscr{R}$-test tries to enforce consistency between $\mathscr{D}$ and $D$: our results are reported in table~\ref{tab:Rtest} for the actual number of events of Xe100. In all scenarios, there is inconclusive evidence against the hypothesis of compatibility between Xe100 and $D$. This can be understood as follows. This test deems the joint evidence in order to make compatible data that come from different regions of the parameter space. The joint evidence $p(\mathscr{D},D)$ is nicely unimodal and sharply peaked around 7 GeV in the DM mass parameter with cross-section that depends on the particle physics scenario. Each of the best fit points are fairly compatible with inference for $D$ alone (see figure~\ref{fig:Combintot}), while individually $\mathscr{D}$ has a very broad and flat posterior probability distribution. However in order to find a common ground the combined set $D$ and the Xe100 data need to tune the astrophysical parameters: apart from the inelastic model (which is fine as it is, as shown already by the $\mathscr{L}$-test) the preferred local circular velocity is now 253 ${\rm km/s}$, with an escape velocity of 568 ${\rm km/s}$ and a DM density at the solar position of $\sim 0.5\, {\rm GeV/cm^{3}}$, values different from the one in table~\ref{tab:astro}. Those values are in the tail of the distribution of the observed values, as given in table~\ref{tab:prior3}. Because of the adjustment of the astrophysical parameters and the widespread original likelihood of Xe100, this test is inconclusive. It is interesting however that the astrophysics in this case plays a fundamental role. Possibly more sophisticated DM halo models, besides the smooth and spherically symmetric ones, may increase the consistency between data sets.

These tests can be easily performed for every exclusion bound versus the combined set, taking into account the time consuming numerical calculations. They are better suited for quantifying consistency between data sets that a global $\chi^2$, because definitely the distribution of the test statistics for detection limits does not certainly follow a $\chi^2$ distribution.

\subsection{Model comparison}\label{sec:modcomp}

The $\mathscr{L}$-test indicates in general an inconsistency between the Xe100 exclusion limit and the combined set $D$, with a statistical significance that depends on the particle physics model $\mathcal{M}_i$. To answer then to the second question addressed in this paper, what is the best particle physics model that can account for the data, we consider only the detection regions, individually and combined together.

\begin{figure}[t]
\includegraphics[width=1.\columnwidth, trim=32mm 60mm 30mm 60mm, clip]{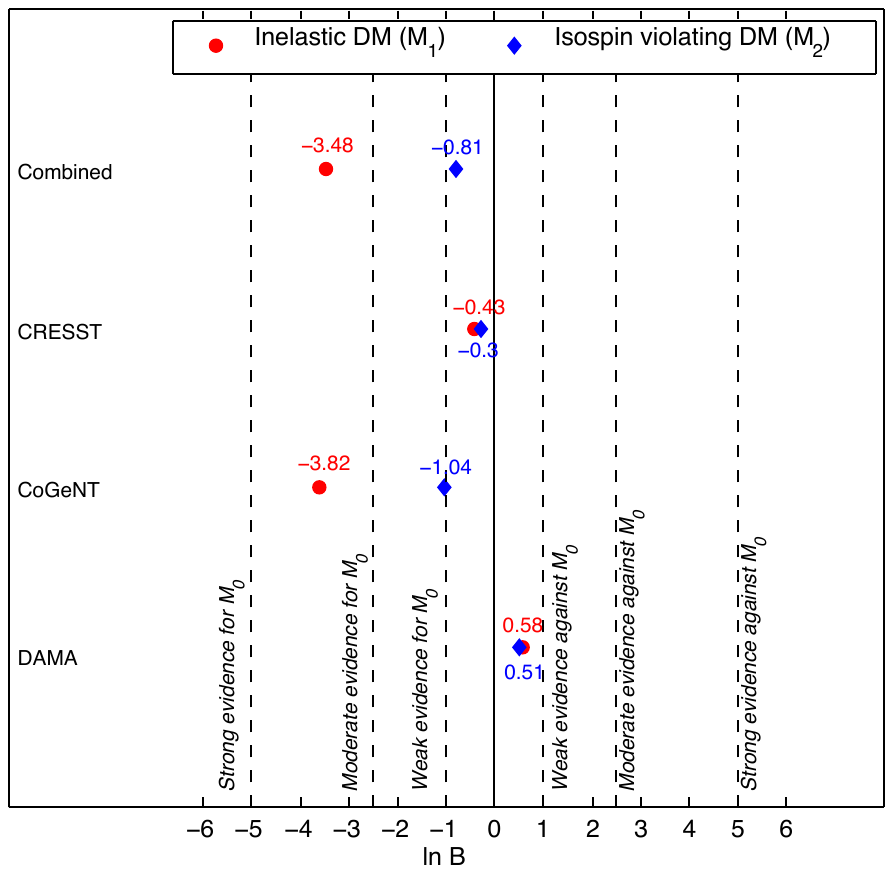}
\caption{Bayes factors for the particle physics scenarios analysed in this work. The experiments are specified on the vertical axis, 
while the different symbols refer to the model for which the Bayes factors have been computed, as labelled in the plot. The numerical value is specified near the data point. The Bayes factors have uncertainties of $(0.02,0.03)$ for the individual experiments and $\sim 0.07$ for the combined analysis. Following Jeffrey's scale in table~\ref{tab:jef}, the vertical lines separate the distinct empirical gradings of the strength of the evidence.
\label{fig:lnB}}
\end{figure}

The main results for Bayesian model comparison are the Bayes factors for the nested models $\mathcal{M}_1$ (inelastic) and $\mathcal{M}_2$ (isospin violating) versus $\mathcal{M}_0$ (elastic). These are shown in figure~\ref{fig:lnB}, while in table~\ref{tab:odds} the corresponding odds against the simplest model are listed, together with the $\Delta \chi^2_{\rm eff}$ and the $p-$values. We recall that both $\mathcal{M}_1$ and $\mathcal{M}_2$ have one extra free parameter with respect to $\mathcal{M}_0$, $\delta$ and $f_n/f_p$ respectively. Astrophysical uncertainties have been marginalised over.

We confirm that for nested models the Bayes factor depends only on the prior of the additional parameter, while the ones related to common parameters cancel out. Indeed in table~\ref{tab:lnBsmh} the Bayes factors for fixed astrophysics are shown: they provide strength of evidence alike figure~\ref{fig:lnB}, where all nuisance and astrophysical parameters are marginalised over. 

\begin{table}[t!]
\caption{Bayes factors for the particle physics scenarios analysed in this work for fixed astrophysical parameters (SMH), for the individual detection regions and for the combined fit. \label{tab:lnBsmh}}
\begin{center}
\begin{ruledtabular}
\begin{tabular}{c|lll}
 & \multicolumn{2}{c}{$\ln B$} \\
Experiments & $\mathcal{M}_1:\mathcal{M}_0$& $\mathcal{M}_2:\mathcal{M}_0$  \\
 \hline
 DAMA & $+0.45$ & $-0.27$ \\
CoGeNT & $-2.52$  & $-0.13$  \\
CRESST & $-0.58$ & $-0.27$   \\
Combined & $-2.38$  & $-0.7$  \\
 \end{tabular}
 \end{ruledtabular}
\end{center}
\end{table}

\begin{table}[t!]
\caption{Odds, $\Delta \chi^2_{\rm eff}$ values and corresponding classical $p$-values of the null hypothesis  for the different particle physics scenarios relative to $\mathcal{M}_0$, elastic SI interaction. We consider a NFW density profile for the DM and marginalise over the astrophysical uncertainties and experimental systematics. The classical $p$-values are obtained via Chernoff's theorem with one extra parameter in the alternative hypothesis relative to the null. \label{tab:odds}}
\begin{center}
\begin{ruledtabular}
\begin{tabular}{c|llll}
& \multicolumn{3}{c}{$\mathcal{M}_i: \mathcal{M}_0$} \\
$\mathcal{M}_1$ Inelastic DM &  odds &  $\Delta\chi^2_{\rm eff}$ &  $p$-values\\
 \hline
 DAMA & $2:1$ & $1.95$  & $0.08$ \\
CoGeNT & $1:37$  & $0.87$ & $0.18$  \\
CRESST & $1:2$ & $0.04$ & $0.42$  \\
Combined & $1:32$  & $0.71$& $0.20$  \\
\hline
$\mathcal{M}_2$ Isospin violating DM &    &   & \\
  DAMA & $2:1$ & $1.88$  & $0.09$\\
CoGeNT & $1:3$  & $0.12$  & $0.36$  \\
CRESST & $1:1$  & $0.03$  & $0.43$ \\
Combined & $1:2$  & $8.56$ & $0.002$  \\
 \end{tabular}
 \end{ruledtabular}
\end{center}
\end{table}

\begin{figure*}[t]
\includegraphics[width=0.325\textwidth,trim=13mm 28mm 10mm 12mm, clip]{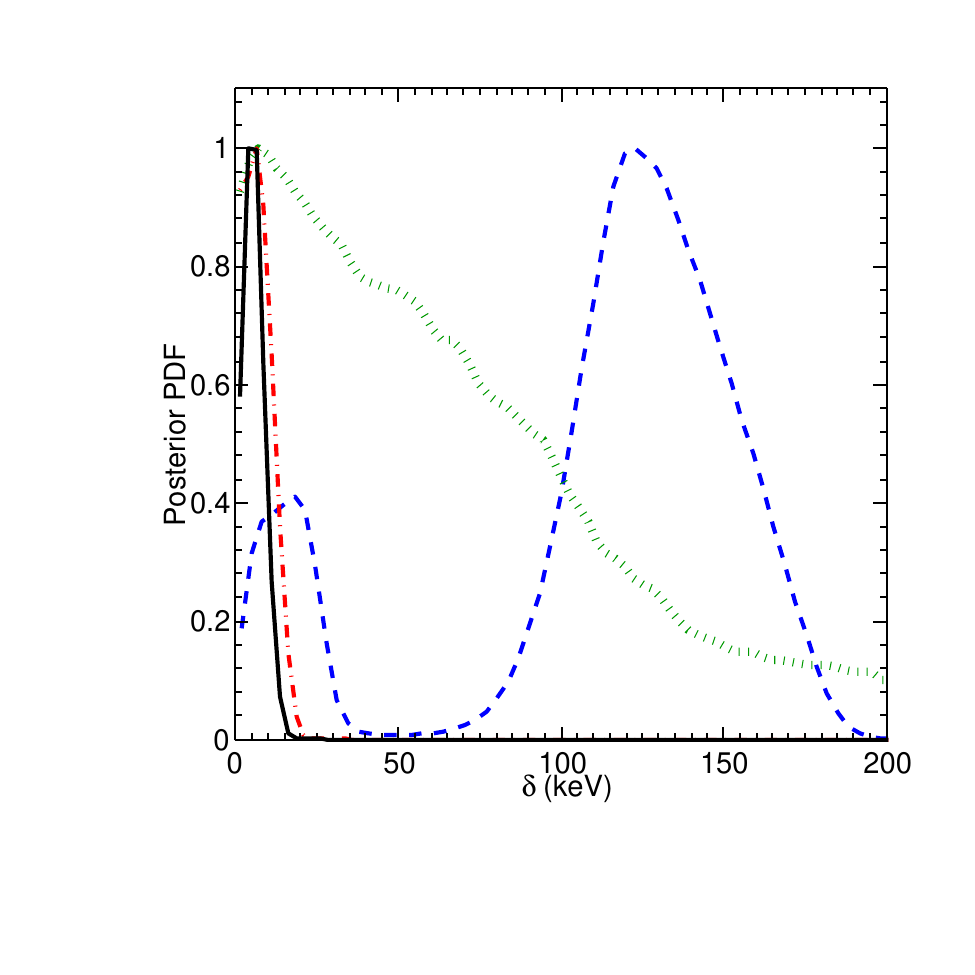}
\includegraphics[width=0.325\textwidth,trim=13mm 28mm 10mm 12mm, clip]{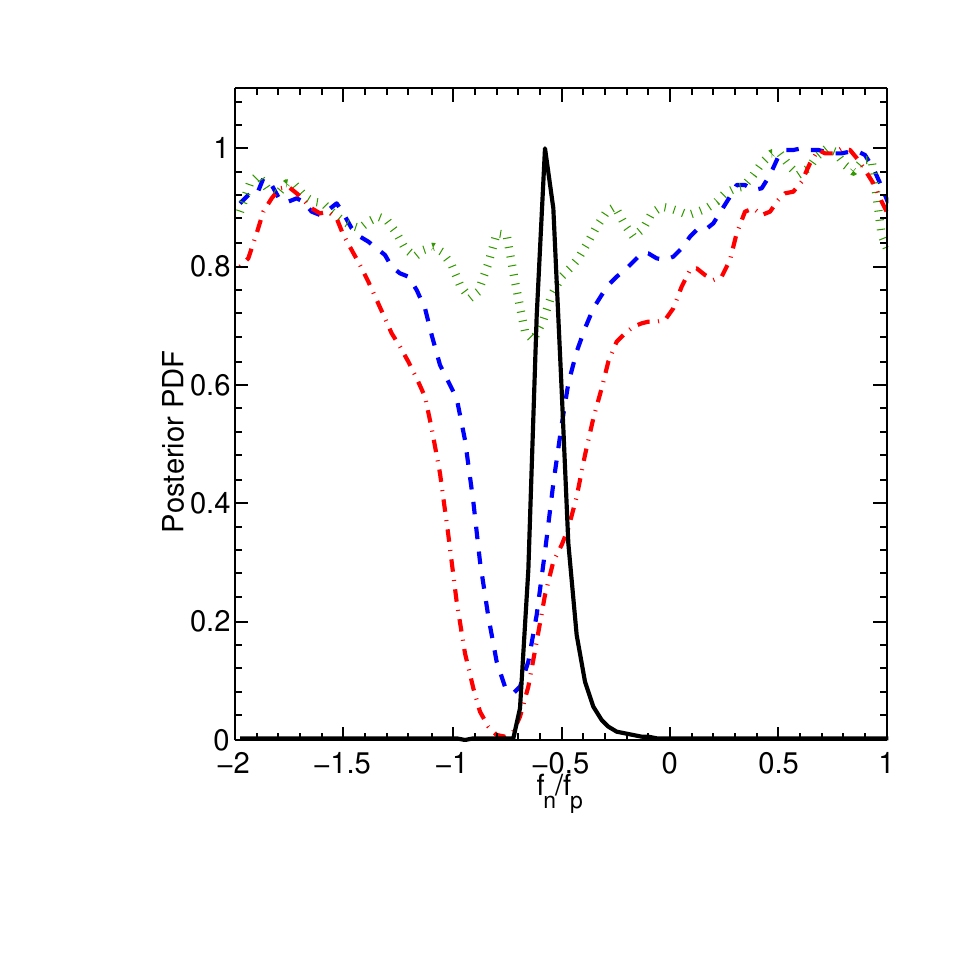}
\includegraphics[width=0.325\textwidth,trim=13mm 28mm 10mm 12mm, clip]{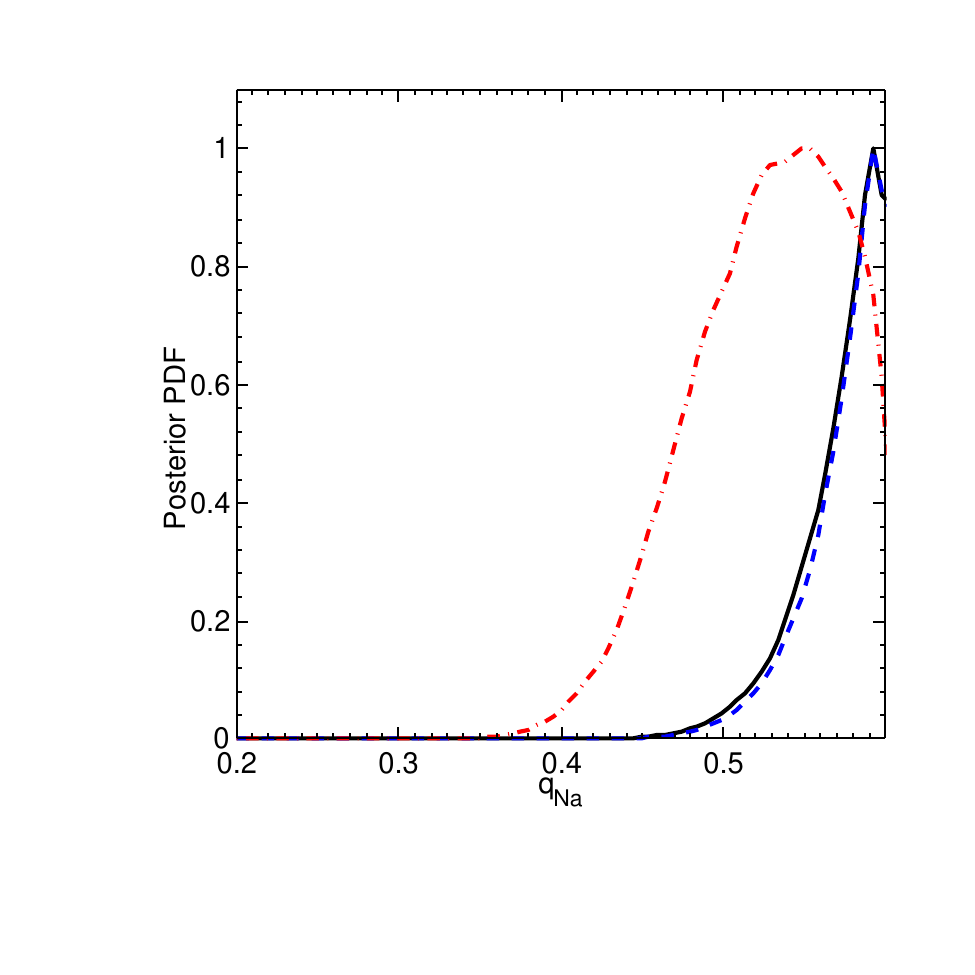}
\caption{{\it Left}: 1D marginal posterior pdf for the mass splitting $\delta$ in the inelastic DM scenario for DAMA (blue dashed), for CoGeNT (red dot-dashed), for CRESST (green dotted) and for the combined fit (black solid). {\it Central:} Same as left for the isospin parameter $f_n/f_p$ in $\mathcal{M}_2$. {\it Right:} 1D marginal posterior for the sodium quenching factor $q_{\rm Na}$ for the combined fit in all the three particle physics scenarios as follows. The black solid line denotes the elastic SI model, the blue dashed the inelastic SI model and the red dot-dashed the isospin violating DM scenario. In all panels the astrophysical uncertainties have been marginalised over.\label{fig:Combin}}
\end{figure*}

From figure~\ref{fig:lnB}, DAMA is the only experiment which shows a positive $\ln B$ for both $\mathcal{M}_1$ and $\mathcal{M}_2$: these scenarios are favoured with respect to the elastic SI model, even though the evidence is inconclusive in both cases, with the odds of only $2:1$ in favour of the most complicated models. This is confirmed by the small values of $\Delta\chi^2_{\rm eff}$, meaning that the additional parameter ($\delta$ or $f_n/f_p$) does not actually improuve the quality of the fit. Regarding these parameters, from figure~\ref{fig:Combin}, we see that the marginal 1D posterior pdf (blue dashed left panel) for $\delta$ has two peaks, one for Na and one for I, while the 1D posterior pdf for $f_n/f_p$ denotes a suppression of the interaction for $-1 < f_n/f_p < -0.5$ (blue dashed central panel) .

On the contrary, CoGeNT prefers the simple elastic scenario, with weak evidence against $\mathcal{M}_2$ and a moderate evidence against $\mathcal{M}_1$. In particular, inelastic SI scattering is disfavoured with the odds of $1:37$ because a large portion of the additional parameter space is wasted and the likelihood does not reach enough improvement not to be deemed by the unpredictive prior. CoGeNT clearly likes light WIMPs with almost elastic collisions (the preferred value for $\delta$ is 6 keV) as confirmed by the 1D marginal posterior pdf in figure~\ref{fig:Combin} (left panel red line). We see an example of Occams' razor principle at work: the more complicated model is disfavoured because the likelihood is not predictive enough to compensate the volume increase due to the extra additional parameter. Less conclusive is the outcome for the isospin violating model with the odds of $1:3$ against $\mathcal{M}_2$, supported by an almost flat $f_n/f_p$  in all prior range except for a deep around $f_n/f_p \sim -0.7$, figure~\ref{fig:Combin} (central panel). 

CRESST indicates inconclusive evidence against both $\mathcal{M}_1$ and $\mathcal{M}_2$. The CRESST data are not able to constrain the nested models with respect to the null hypothesis, the odds are at most $1:2$. This is confirmed by the broad 1D marginal posterior pdf for both $\delta$ and $f_n//f_p$ in figure~\ref{fig:Combin} (left and central panels, green dotted lines). The behavior of $f_n/f_p$ is a consequence of multi-target detectors: for instance depending on the atomic element, different values of $f_n/f_p$ might be suppressed, leading in complex to an  almost flat behavior.

\begin{figure}[t]
\includegraphics[width=1.\columnwidth,trim=25mm 65mm 25mm 65mm, clip]{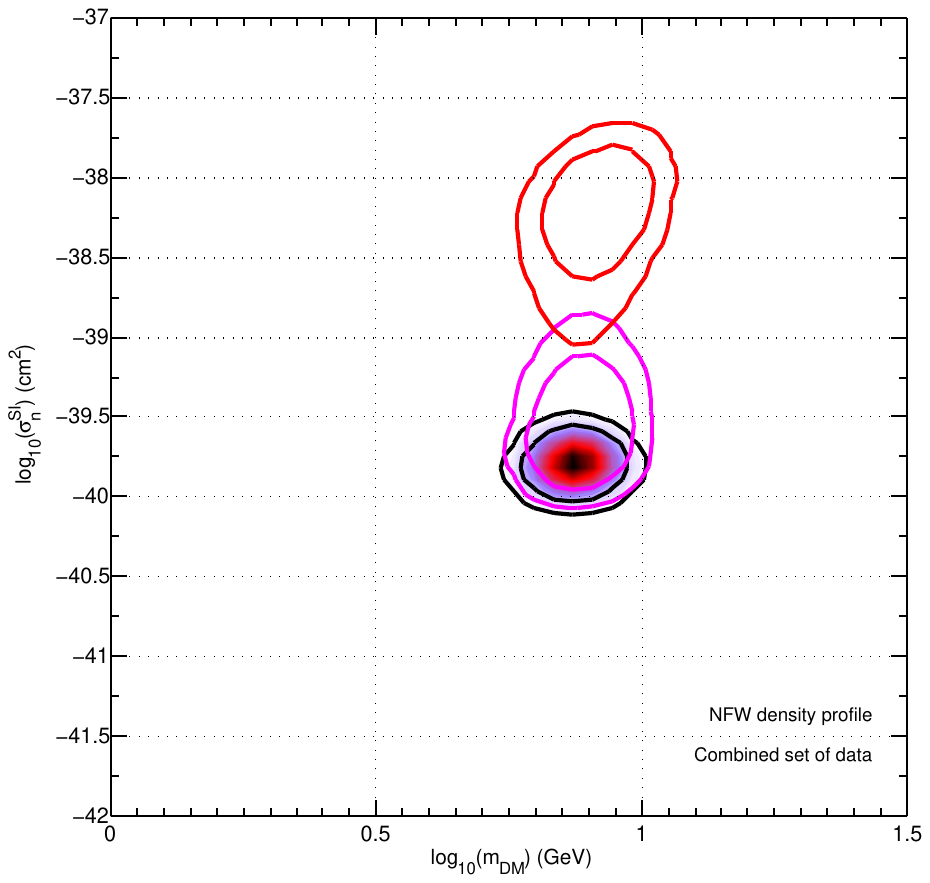}
\caption{2D credible regions  at 90\% and 99\% for the combined data set ($D = \{{\rm DAMA, CRESST, CoGeNT}\}$) for the three particle physics scenarios considered in this work, combined in a single plot. Elastic SI interaction is given by the shaded region, the magenta contours (lower non filled region) are for the inelastic scenario, while the red curves (upper non filled region) are for the isospin violating model. All the nuisance parameters have been marginalised over, as well as the additional model parameters.\label{fig:Combintot}}
\end{figure}

The outcome of model selection for the combined fit is driven by CoGeNT data: indeed $\ln B$ indicates a moderate evidence against $\mathcal{M}_1$ with the corresponding odds of $1:32$. The combined posterior pdf (black solid) follows closely the one of CoGeNT (red dot-dashed) in the left panel in figure~\ref{fig:Combin}. The 90\% and 99\% credible regions in the \{$m_{\rm DM}, \sigma_n^{\rm SI}$\}-plane are shown in figure~\ref{fig:Combintot} (magenta non filled). The inelastic SI scenario favours similar values for mass and cross-section as elastic case (shaded region), that is $m_{\rm DM} \sim 7$ GeV and $\sigma_n^{SI} \sim 10^{-40} {\rm cm^2}$. One has to look along the third direction to check if the agreement provides really a good fit to all of the experiments: regarding the astrophysical parameters the preferred values are  $\rho_{\odot} = 0.34\,  {\rm GeV/cm}^{3}$, $v_0 = 212\, {\rm km/s}$, $v_{\rm esc} = 556\,  {\rm km/s}$ for $\mathcal{M}_0$ and $v_0 = 220\,{\rm km/s}$ and $\rho_{\odot} = 0.37\,{\rm GeV/cm}$ for $\mathcal{M}_1$. This long list of preferred values demonstrates that the nuisance parameters select values which are in line with the best fit point of the individual experiments. The only exception is the Na quenching factor, right panel of figure~\ref{fig:Combin}: it peaks at $\sim 0.55$ for all the particle physics models. Even though $q_{\rm Na}$ tends towards a corner of the prior range, this value is still compatible with the experimental allowed range~\cite{Tretyak:2009sr,*Bernabei:1996vj,*Smith:1996fu,*Fushimi:1993nq,*Chagani:2008in}.

On the contrary of inelastic SI scattering, the evidence against $\mathcal{M}_2$ is only inconclusive. A frequentist approach would have preferred this model with respect to elastic SI interaction on the line with the `chi by eye' outcome (as we discussed for figure~\ref{fig:All_NFW}). The $p$-value is 0.002 corresponding to 3$\sigma$ against the null, having considered a gaussian distribution for the test statistic. This is an example of Lindley's paradox (namely Bayesian model selection returning a different result from classical hypothesis testing, see~\cite{Trotta:2005ar} and references therein): looking at figure~\ref{fig:Combin}, second panel, the 1D posterior pdf for $f_n/f_p$ is sharply peaked around its preferred value, meaning that the broad range prior is diluting the evidence for $\mathcal{M}_2$, contrary to the single experiments, where $f_n/f_p$ is non negligible in all the prior range. The marginal 2D posterior pdf in the $\{m_{\rm DM}, \sigma^{\rm SI}_n\}$-plane is given by the red contours in figure~\ref{fig:Combintot}, and prefers large values of the cross-section for a 10 GeV DM mass with respect to the other scenarios. Again the astrophysical parameters are in line with those of the single experiments.

Resuming, we argue that the current experimental situation disfavours the inelastic DM picture because of CoGeNT data. The $p$-value of 0.2 corresponds formally to a 1.3 $\sigma$ exclusion with respect to the null hypothesis. On the other hand the outcome between elastic and isospin violating SI scattering has an inconclusive strength of evidence, meaning that the complexity due to the extra free parameter is not supported yet. 

Our conclusions are robust against changes in prior range of the extra free parameter. By means of the SDDR we evaluate the impact of changing the prior range of the extra free parameter. The odds for a more complex model can be made arbitrarily small by increasing the width of the priors on the additional parameters or by choosing uniform priors on non-linear functions of this parameter. Note that a rescaling by a factor of 2 ($\delta: 0 \to 100$ keV instead of $0 \to 200$ keV) would still disfavour moderately $\mathcal{M}_1$ with respect to $\mathcal{M}_0$ for CoGeNT. On the other hand it can turn it into a positive evidence for $\mathcal{M}_1$ versus $\mathcal{M}_0$ for DAMA and CRESST, although still inconclusive. The main conclusion for the combined set would still be valid as well. For isospin violating model, a reduction in the prior range by a factor of $\lambda=2$ would still lead to inconclusive evidence between $\mathcal{M}_2$ and $\mathcal{M}_0$ in all experiments.

\section{Conclusions}\label{sec:concl}

Currently the direct detection experiments exhibit contrasting outcomes, leading to an ambiguous situation. We have applied Bayesian statistical tools to three model independent scenarios for spin-independent scattering: elastic, inelastic and isospin violating. We have resumed the state of the art of these three models
using the latest results of DAMA, CoGeNT, CRESST, Xe100, KIMS and bubble chamber experiments; the experimental systematic have been carefully modelled in the likelihood.  We argued that the usual `chi by eye' consistency test may induce to misleading interpretation of consistency between data sets in certain cases.

We therefore have rigorously quantified the tension between detection regions at low DM mass (data set $D$) and Xe100 exclusion bound (data set $\mathscr{D}$), by means of Bayesian statistical techniques. Using Bayesian evidence we have performed two statistical tests that look for inconsistency between data sets and the underlying WIMP theoretical model. The model comparison test, or $\mathscr{R}$-test leads to inconclusive result, while the predictive likelihood test has a striking outcome. We have found that the inelastic SI scenario is the favoured one under the hypothesis of a global explanation of both Xe100 and the combined set. The same data sets appear to be inconsistent in both elastic and isospin violating models with a significance at $3\sigma$ and $2\sigma$ level respectively, if a reasonable hypothesis on the observed number of events in the Xe100 detector is made. Notice that the DM halo distribution plays an important role for the joint set $\{\mathscr{D},D\}$: the data adjust the values of the astrophysical parameters to find a common ground of agreement. The interpretation can be twofold: either one has to look for experimental systematics and/or astrophysical modelling that could accommodate both $\{\mathscr{D},D\}$ either the discrepancy can be seen as an evidence against the DM explanation of current data.

Considering only the detection regions, we have performed Bayesian model selection to single out the best particle physics scenario that phenomenologically accommodates the data sets of DAMA, CoGeNT and CRESST individually and in a combined fit. It turns out that the isospin violating picture has odds similar to the simplest elastic SI interaction: the extra parameter $f_n/f_p$ is not supported by the current data. The inelastic SI model is disfavoured with the odds of $1:32$ with respect elastic scattering because it does not provide a good fit for CoGeNT, namely it is penalised because of the unpredictive broad prior.

We remark that Bayesian model comparison outcomes point somehow towards the opposite direction than the consistency picture between Xe100 and the combined set. In other words the situation is still too tangled to draw a conclusive answer; more data are needed as well as public likelihoods given by the collaboration in order to properly take into account the experimental systematics.

\section*{Acknowledgments} 

It is a pleasure to thank N. Fornengo, C. Ringeval and R. Trotta  for very helpful discussions as well as the Cosmo computing resource at CP3 of Louvain University for making possible the numerical analyses. The author is partially supported by a European Research Council Starting Grant, under grant agreement No. 277591, PI G. Bertone.

\appendix
\section{Details on experimental likelihoods}\label{sec:appa}

\paragraph*{XENON100} 
This experiment is currently running at Laboratori Nazionali del Gran Sasso in Italy. It has recently released the scientific run based on 224.6 live days of data taking with a fiducial volume for the detector of 34 kg~\cite{:2012nq}. The blind analysis, after cuts optimized for DM searches, has reported 2 candidate events for WIMP recoils ($N_{\rm obs}=2$) with an expected background of 1 event (more precisely the background with its uncertainty is $\bar{B} \pm \sigma_B =1 \pm 0.2$). After cuts the total exposure is equivalent to 2323.7 kg days, value used in this analysis. The likelihood $\ln \mathcal{L}_{\rm Xe100}$ is the same as in~\cite{Arina:2011si}, with updated total exposure and number of observed events, and receives contribution from two parts:
\begin{enumerate}
\item $\ln \mathcal{L}_{\rm events}$ is the Poisson probability distribution for having seen 2 events with a background of 1 event. In this analysis we marginalise over the background analytically:
\begin{eqnarray}
& \ln \mathcal{L}_{\rm events}(N_{\rm obs}| S,B)  =   - S-\bar{B}+ \frac{\sigma_B^2}{2}+2 \nonumber\\
& +  \ln \left[\frac{ \sigma_B^2 + (S + \bar{B} - \sigma_B^2)^2}{4} \right]  \,;
\end{eqnarray}
\item $\ln\mathcal{L}_{\rm L_{\rm eff}}$ is a Gaussian distribution function that models the uncertainty under threshold of $L_{\rm eff}$, which is the conversion factor between nuclear recoil energy $E$ and photo-electron (PE) produced in the primary scintillation light ($S_1$ signal). The actual nuisance parameter is called $m$.
\end{enumerate}
The detection range for DM in the $S_1$ variable is $3 \to 20$ PE, contrary to the old run which used $4 \to 30$ PE~\cite{Aprile:2011hi}.  As already remarked in~\cite{Arina:2011si}, our likelihood is an approximation of the one provided by the XENON100 collaboration in~\cite{Aprile:2011hx}, because the spectral informations are not available. The $90_S\%$ confidence level in the plane $\{m_{\rm DM},\sigma_n^{SI}\}$ corresponds to $\Delta \chi_{\rm eff}^2 \leq 3.1$.

\paragraph*{CRESST} The Cryogenic Rare Event Search with Superconducting Thermometers experiment is located at the Laboratori Nazionali del Gran Sasso in Italy. The detectors are scintillators made by $\rm CaWO_4$ crystals. The latest release covers the period between July 2009 and March 2011 and collects the data from eight detector modules for a total exposure after cuts of 730 kg days. The analysis pursued by the collaboration counts 67 events ($N_{\rm obs}$), which can not be all accounted for by known background, leading to a hint of detection with a statistical significance of more than $4 \sigma$~\cite{Angloher:2011uu}.

The discrimination between background and nuclear recoil is obtained by the interplay of the phonon channel and the scintillation signal. The phonon signal provides a measurement of the total energy deposited by the interaction, while the scintillation channel serves to discriminate the type of interaction (different particles give a different light yield). However this information is not provided by the collaboration. We construct then an approximate likelihood based on the total number of events in each module plus the total spectral information~\cite{Angloher:2011uu}. We suppose that all detector modules have the same total exposure, that is 730/8 kg days. The typical energy range for DM searches is 12-40 keV, however each detector module has is own energy threshold, as detailed in table 1 of~\cite{Angloher:2011uu} together with the total number of events observed in each module. 

The first part of the likelihood models the total number of events seen in each detector module and has the form:
 \begin{equation}
 \ln\mathcal{L}_{\rm module} = \sum_{i=1}^8 \ln\mathcal{L}_i(n_{\rm obs}^i | S_i, \sum_j B_{ij}) \,,
 \end{equation}
 where the sum runs over all detector modules. In each detector the likelihood is given by the Poisson probability of observing $n_{\rm obs}^i$ events for a given WIMP signal $S$ and a given background $B_i = B_{i\alpha} + B_{i\, e/\gamma} + B_{i\, n} + B_{i\, \rm Pb}$:
 \begin{eqnarray}
&  \ln\mathcal{L}_i (n_{\rm obs}^i | S_i, \sum_j B_{ij}) =\nonumber\\
& \ln \left[\frac{(S_i+ \sum_j B_{ij})^{n_{\rm obs}^i} \, \exp\left(-S_i-\sum_j B_{ij}\right)}{n_{\rm obs}^i !}\right]\,.
  \end{eqnarray}
The index $j$ runs over the 4 different sources of background defined above, while $i=1,...,8$ denotes the modules. The second part of the likelihood, $\ln\mathcal{L}_{\rm Spectral}$, is modelled with a Poisson distribution as well and uses the spectral information given in figure 5 of ~\cite{Angloher:2011uu}. Each bin has a width of 1 keV and the energy ranges from 10 to 40 keV, for a total of 30 bins.

The identified background sources are:
\begin{enumerate}
 \item Leakage of $e/\gamma$ at low energies, as a total of 8 events ($B_{e\gamma}$);
 \item Scattering from $\alpha$ particles, due to the overlap of the alpha recoil band with the acceptance region ($B_{\alpha}$);
 \item Pb recoils due to alpha decay of Polonium at energy around 130 keV ($B_{\rm Pb}$);
 \item Neutron scatterings off Oxygen mainly ($B_n$).
 \end{enumerate} 
The background is a source of systematics and should be marginalised over to obtain the credible regions in the $\{m_{\rm DM}, \sigma_n^{SI}\}$-plane. The $e/\gamma$ background is not varied and we suppose that in the first energy bin of each module it contributes with one event. The $\alpha$ background has constant rate in each energy bin and is described by the total number of observed $\alpha$ events  such that:
\begin{equation}
N_\alpha = \sum_{i=1}^8 B_{i \alpha}\,.
\end{equation}
 The contamination due to Pb decay is parametrized as equation 1 of~\cite{Angloher:2011uu}:
 \begin{equation}
 \frac{{\rm d}B_{\rm Pb}}{{\rm d}E} = C_{\rm Pb}  \left[ 0.13 + \exp\left( \frac{E-90}{ 13.72} \right) \right]\,,
 \end{equation}
 with the normalization $C_{\rm Pb}$ let free to vary. Finally the neutron background is parameterized following equation 10 in~\cite{Angloher:2011uu}, with a free normalization $N_n$:
 \begin{equation}
 B_n = N_n \left[\exp\left(-\frac{E_{\rm min}}{23.54}\right)-\exp\left(-\frac{E_{\rm max}}{23.54}\right)\right]\,.
 \end{equation}
 where $E_{\rm min,max}$ are the extreme of each energy bin/range.
 
The total likelihood is then:
\begin{equation}
\ln\mathcal{L}_{\rm CRESST}(N_{\rm tot}|S,B) = \ln\mathcal{L}_{\rm module} +   \ln\mathcal{L}_{\rm Spectral} + \ln\mathcal{L}_{B}\,,
\end{equation} 
and depends on the three nuisance parameters from background modelling, resumed in table~\ref{tab:prior1}. For each nuisance parameter we use a Gaussian prior centered on the preferred value, as indicated by the collaboration: $\bar{B}_\alpha \pm \sigma_\alpha = 9.2 \pm 2.3$, $\bar{B}_n \pm \sigma_n = 9.7 \pm 5.1$ and $\bar{B}_{\rm Pb} \pm \sigma_{\rm Pb} = 19\pm 5$. The sum of the Gaussian distributions gives $\ln\mathcal{L}_{B}$. Note that the reported energies are already the bolometric ones: we will not be able to fold into the Bayesian analysis the uncertainties related to the quenching factors. Indeed these have been used by the collaboration to define the acceptance region in each detector module and for each target nucleus.

The CRESST commissioning run on W~\cite{Angloher:2008jj,Brown:2011dp,Kopp:2011yr} is constraining part of the parameter space of the CRESST-II run, in particular the region at relatively high DM mass. We do not however consider it since other bounds will reveal to be more stringent.
 
 \begin{figure*}[t]
\includegraphics[width=0.32\textwidth,trim=14mm 70mm 23mm 75mm, clip]{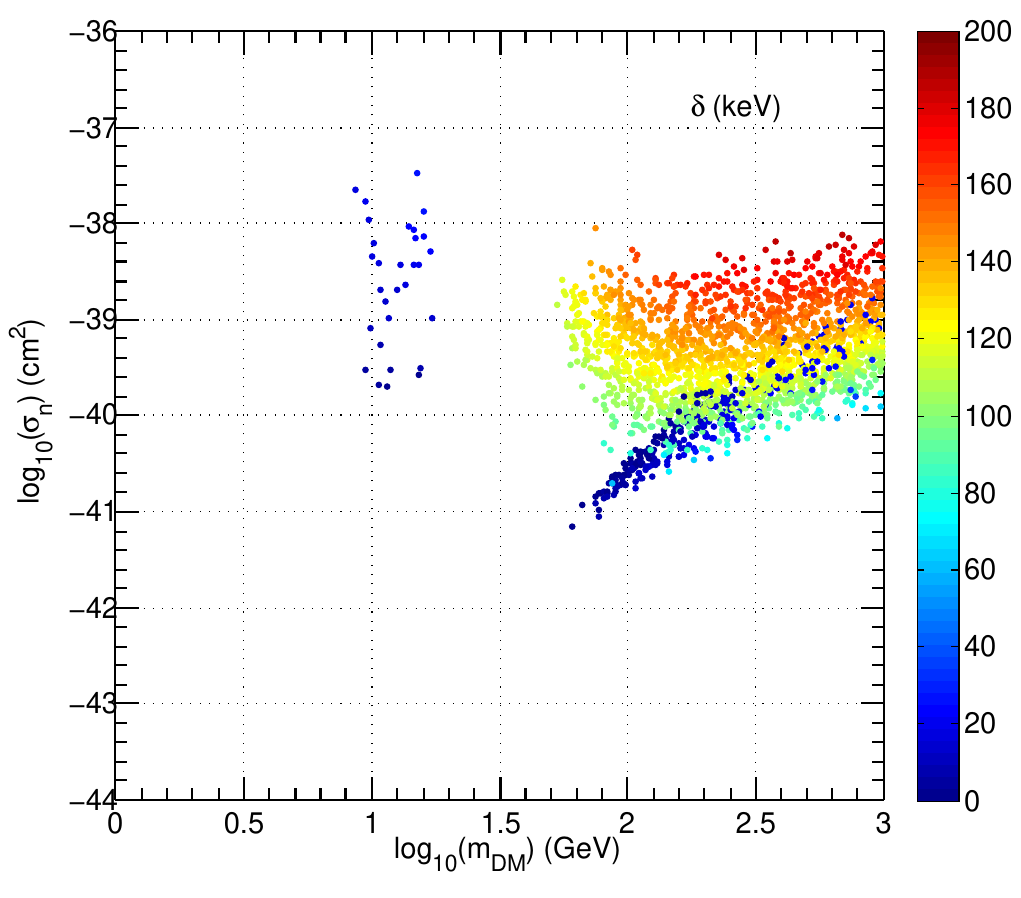}
\includegraphics[width=0.32\textwidth,trim=14mm 70mm 23mm 75mm, clip]{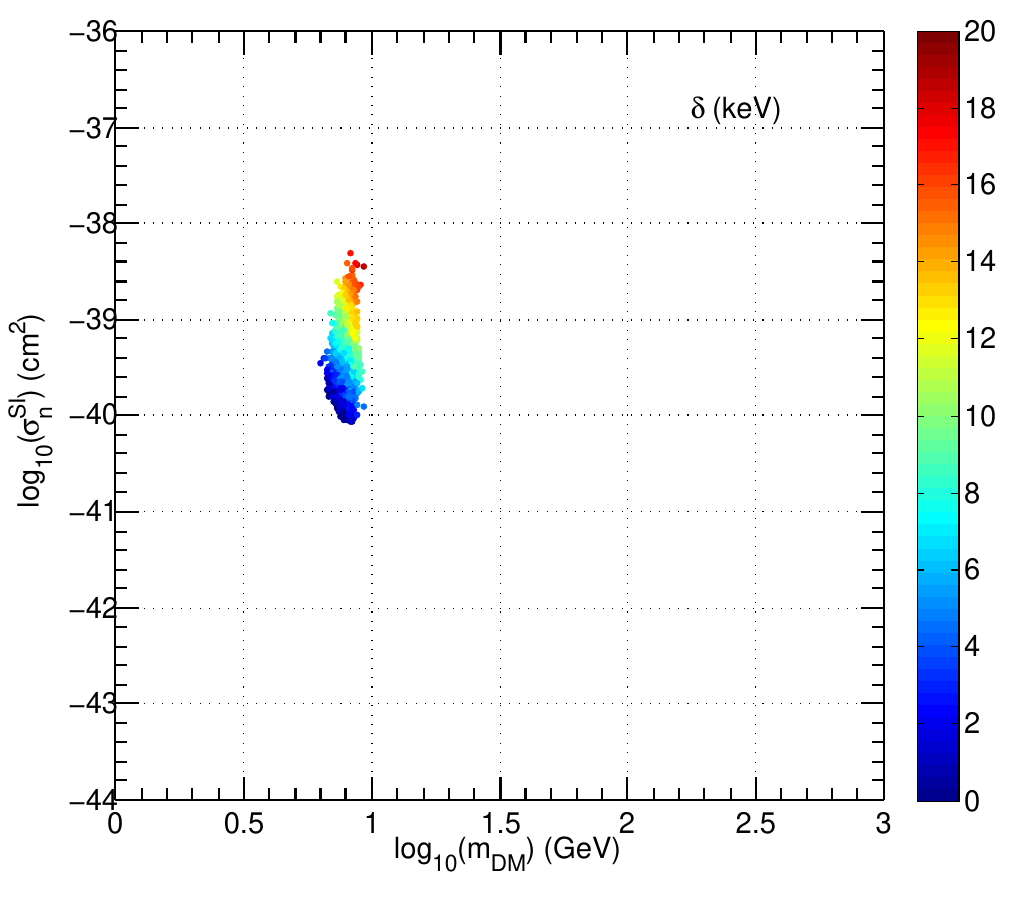}
\includegraphics[width=0.32\textwidth,trim=14mm 70mm 23mm 75mm, clip]{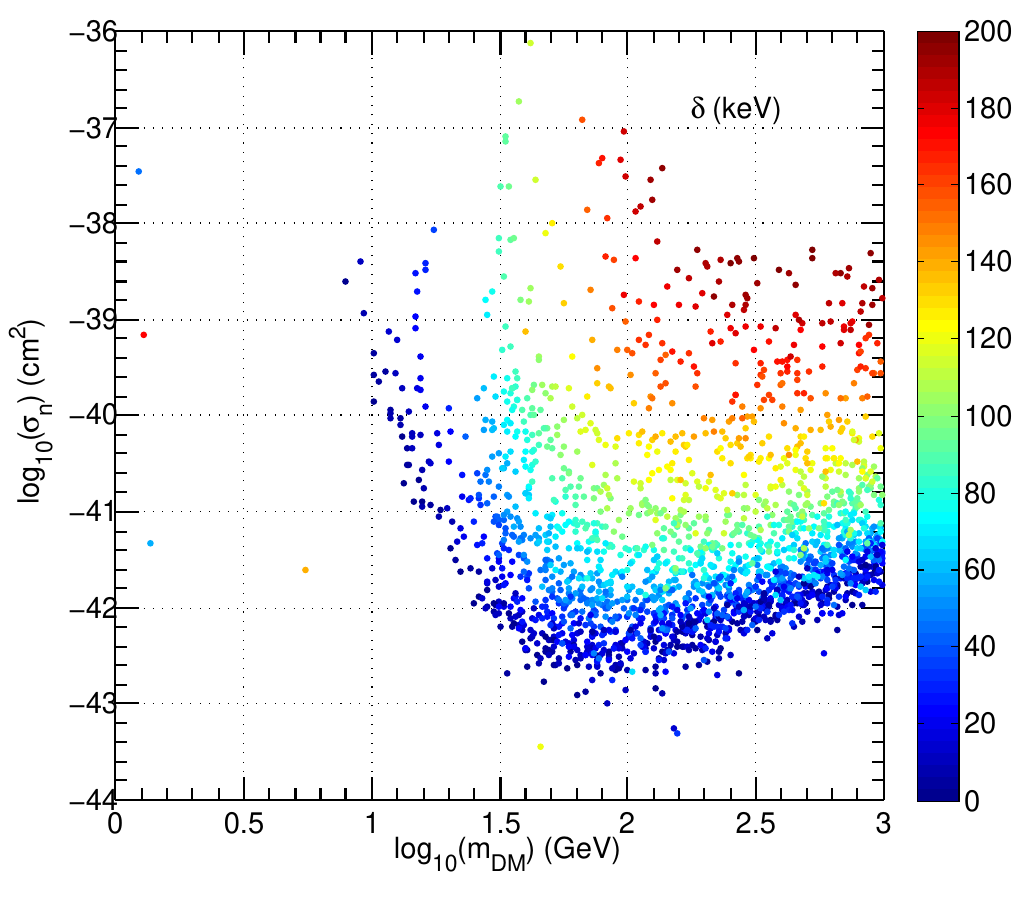}
\caption{Inelastic scattering parameter $\delta$. {\it Left}: 3D marginal posterior pdf for \{$m_{\rm DM},\sigma_n^{\rm SI},\delta$\}, where the $\delta$ direction is represented by the colour code, for DAMA. 
{\it Central and right}:  Same as left for CoGeNT and CRESST respectively. The astrophysical nuisance parameters are fixed at their central value (SMH), while all the experimental systematics are marginalised over.
\label{fig:deltaSMH}}
\end{figure*}

 \paragraph*{PICASSO} The experiment~\cite{Archambault:2012pm} is located at SNOLAB, the canadian underground laboratory in the Vale Creighton mine. This search for DM uses superheated liquid droplets, a variant of the bubble chamber technique, with $\rm C_4F_{10}$ as liquid target material. PICASSO has become sensitive to low mass WIMPs, thanks to the lightness of the detector material, to the low energy threshold (around 1.7 keV) and to the total exposure of 114 kg days (on $^{19}F$). It was although originally planned for investigating WIMP spin-dependent interaction, because of its unpaired proton in $^{19}F$. The collaboration has estimated that the scattering off $^{12}C$ contributes by 10\% for SI interaction, which we take into account. 

Cosmic muons, $\gamma$ and $\beta$ particles are well separated as background, while the main contamination comes from neutron and in particular $\alpha$ particles. In our analysis we use the data of figure 5 of~\cite{Archambault:2012pm}, which arise from a combination of all detectors and for which the background has already been subtracted.  We can not therefore take into account the uncertainties due to the $\alpha$ background, however we include in the analysis a $5\%$ of uncertainties from systematics, as quoted by the collaboration. The nuisance parameter $a(T)$ is varied with a flat prior within its measured experimental range, that is from 1 to 11. The likelihood is then defined as:
\begin{equation}
 \ln \mathcal{L}_{\rm PICASSO} = - \frac{\chi^2}{2} - \sum_i \ln \left( 2 \pi \sigma_i^2 \right) \,,
\end{equation}
where the index $i$ runs over the eight data bins and $\sigma_i$ are the corresponding error bars. The last factor is merely a normalization not important for inference however crucial when computing the Bayesian evidence. The $90_S\%$ confidence level in the plane $\{m_{\rm DM},\sigma_n^{SI}\}$ corresponds to $\Delta \chi_{\rm eff}^2 \leq 4.6$.

\paragraph*{SIMPLE-II} The Superheated Instrument for Massive ParticLe Experiments (SIMPLE hereafter) is operating in the Low Noise Underground Laboratory in southern France. It consists of 15 superheated droplets detector of $\rm C_2 Cl F_5$. As in the case of PICASSO experiment, it is well suited to probe the light DM with SI interaction, as well as for constraining the spin-dependent cross-section for the whole WIMP mass range.

We neglect the phase I in~\cite{Felizardo:2010mi} and use the most recent run of 2010, which has an improved neutron shield. The final stage of phase II has been released in~\cite{Felizardo:2011uw} and encompasses few months of data taking. The total exposure after cuts is 6.71 kg days, with one event observed ($N_{\rm obs}=1$) and a neutron background estimated to be $\bar{B}+\sigma_B = 2.2 \pm 0.3$, while the alpha background has been estimated negligible. The likelihood is therefore given by the Poisson probability of observing $N_{\rm obs}$, marginalised analytically over the background, as described in~\cite{Arina:2011si}:
\begin{equation}
\ln \mathcal{L}_{\rm SIMPLE}(N|S) =  - S - \bar{B} + \frac{\sigma^2_B}{2} + \ln \left( S +\bar{B} - \sigma^2_B \right)\,.
\end{equation}
The observed rate is calculated using equation~\ref{eq:rsupheat}, with the parameter $a(T)$ modelled by a Gaussian prior centered on its mean value $4.2$ and with standard deviation of $0.3$. The energy threshold is set to 8 keV. The $90_S\%$ confidence level in the plane $\{m_{\rm DM},\sigma_n^{SI}\}$ corresponds to $\Delta \chi_{\rm eff}^2 \leq 3.27$.

\paragraph*{KIMS} The Korea Invisible Matter Search (KIMS) experiment~\cite{Kim:2012rz} is running at the Yangyang Underground Laboratory in Korea and is made of $\rm CsI(Tl)$ scintillator crystals.  The collaboration has released the data collected from September 2009 to August 2010 for a total exposure of 24524.3 kg days. We construct a Gaussian likelihood based on the counts/keV/kg/day given in figure 4 of~\cite{Kim:2012rz}, which arise from the 8 detectors with the lowest alpha particle contamination. The energy range of the experiment is $3-11$ keVee. The detectors are scintillators, hence the quenching factor of Iodine and Cs are two nuisance parameters, which we vary with a flat prior in the allowed experimental range. In addition a third nuisance parameter comes from the $\alpha$ background, $B_\alpha$, described by a Gaussian distribution centered on $\bar{B}_\alpha \pm \sigma_\alpha = 0.07 \pm 0.02$ counts/keV/kg/day (derived from table I of~\cite{Kim:2012rz}). The $90_S\%$ confidence level in the plane $\{m_{\rm DM},\sigma_n^{SI}\}$ corresponds to $\Delta \chi_{\rm eff}^2 \leq 4.6$.

\section{Details on parameter inferences}\label{sec:appb}

Here we provide an in-depth discussion about the dependence of the detection regions on extra free theoretical parameters and additional details about each individual experiment considered in this work.

\paragraph*{Elastic SI scattering} All the comments below refer to figure~\ref{fig:All_SMH}, and are applicable both to fixed or marginalised astrophysics.

\begin{itemize}
\item DAMA: we remember that the 1D posterior pdf for $q_{\rm Na}$ is flat all along the prior range, given by the measured experimental range~\cite{Tretyak:2009sr,*Bernabei:1996vj,*Smith:1996fu,*Fushimi:1993nq,*Chagani:2008in}. 
\item CoGeNT: marginal posterior is nicely multimodal and the best fit point is at $m_{\rm DM} = 7$ GeV and $\sigma^{SI}_n = 2 \times 10^{-40} {\rm cm^2}$.
\item CRESST: our analysis does not provide a closed region at large WIMP masses, as in~\cite{Angloher:2011uu}, because we could not include the yield information in the likelihood, while we agree with other public analyses, see {\it e.g.}~\cite{Fornengo:2011sz}. The wide region is due to volume effects because of the marginalisation over the background. Since the marginal posterior pdf is highly multimodal inference for the best fit point is meaningless.
\item Xe100: our exclusion limit agrees well with the one provided by the collaboration, despite the marginalisation over $L_{\rm  eff}$. The nuisance parameter $m$ is centered around the best fit measured by the XENON100 collaboration~\cite{Plante:2011hw}\footnote{The latest measurements of $L_{\rm eff}$ by XENON100 has been released very recently~\cite{Aprile:2012an} and shows a flat behavior for $L_{\rm eff}$ below 3 keV. We use~\cite{Plante:2011hw} however for the analysis, as the XENON100 collaboration.}. We attribute the strong constraining power at low WIMP mass to the low threshold of 3 PE.
\item Bubble chambers:  PICASSO is more constraining than SIMPLE at low WIMP mass. As expected both limits become negligible as soon as the DM mass gets larger than $\sim 20$ GeV. We have marginalised over the slope of the threshold temperature $a(T)$, therefore our bounds are less constraining that the one presented by the collaborations. We have although checked that for fixed value of $a(T)$ both limits agree well with~\cite{Archambault:2012pm} and~\cite{Felizardo:2011uw}.
\item CDMSSi: it is competitive with PICASSO and SIMPLE for DM masses below 20 GeV.
\item KIMS: not relevant for this scenario.
\end{itemize}

\begin{figure*}[t]
\includegraphics[width=0.325\textwidth,trim=34mm 90mm 38mm 95mm, clip]{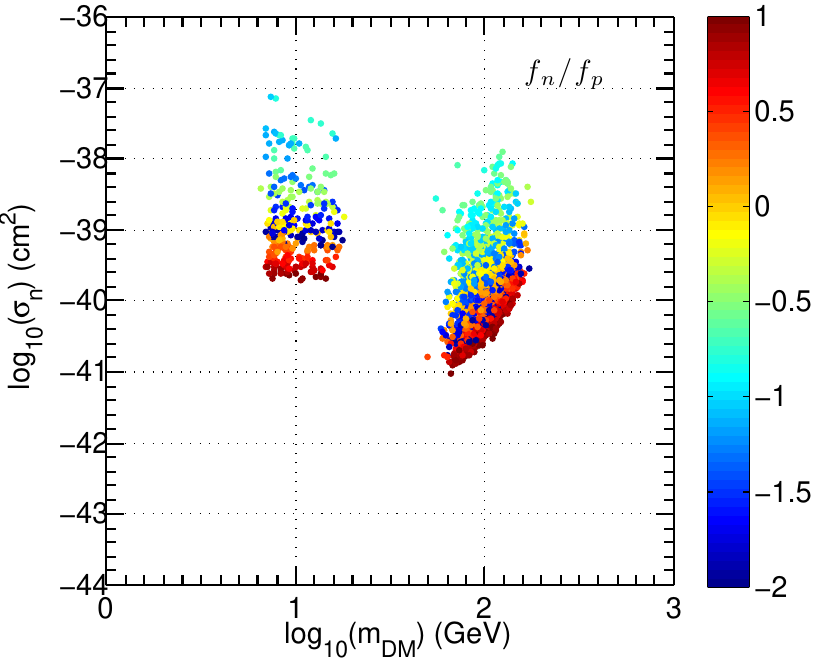}
\includegraphics[width=0.325\textwidth,trim=34mm 90mm 38mm 95mm, clip]{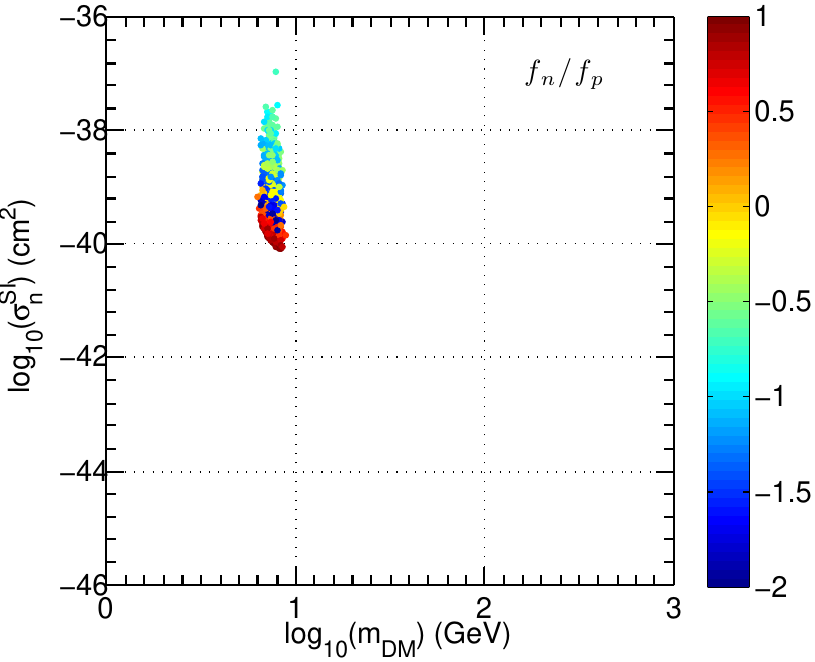}
\includegraphics[width=0.325\textwidth,trim=34mm 90mm 38mm 95mm, clip]{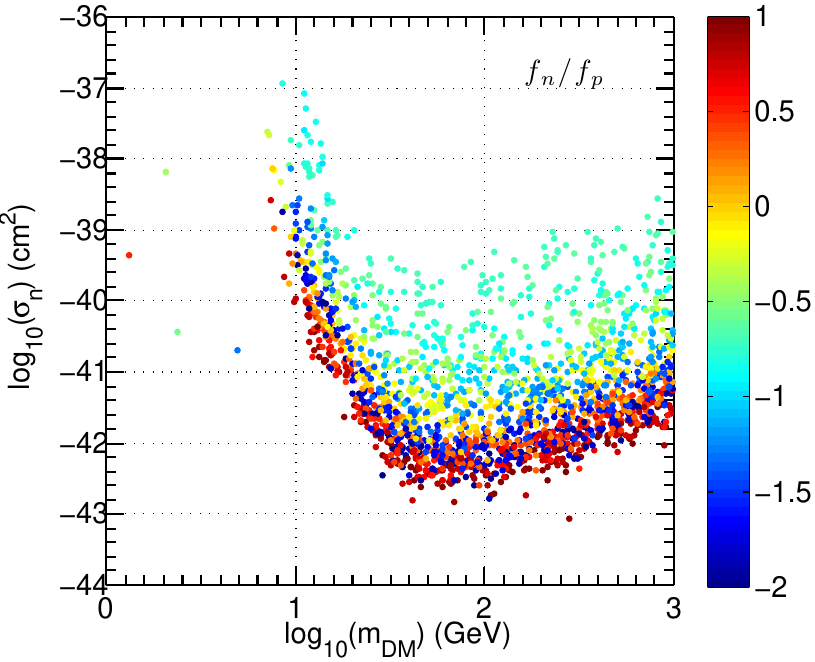}
\caption{Isospin violating parameter $f_n/f_p$. {\it Left}: 3D marginal posterior pdf for \{$m_{\rm DM},\sigma_n^{\rm SI},f_n/f_p$\}, where the $f_n/f_p$ direction is represented by the colour code, for DAMA. {\it Central and right}:  Same as left for CoGeNT and CRESST respectively. The astrophysical nuisance parameters are fixed at their central value (SMH), while all the experimental systematics are marginalised over.
\label{fig:fnfpSMH}}
\end{figure*}

\paragraph*{Inelastic SI scattering} 
The comments below refer to figure~\ref{fig:All_NFW} (left panel) and figure~\ref{fig:deltaSMH}, and are valid both for SMH and marginalised astrophysical case.
\begin{itemize}
\item DAMA: The region at large DM mass is due to scattering off Iodine, while the region at $\sim 10$ GeV is due to scattering off Sodium. The DAMA data are not constraining enough to select a value for the quenching factors, that again has a flat marginal 1D posterior pdf. The parameter $\delta$ has a definite trend, as it is depicted in figure~\ref{fig:deltaSMH} left panel:  for the scattering off Iodine the larger the cross-section the larger the mass splitting is, while the small island due to Sodium interactions allows only small mass splitting of the order $\mathcal{O}(10-20)$ keV. 
\item CoGeNT: the detection region depends only on $\delta < 20$ keV (central panel of figure~\ref{fig:deltaSMH}, note the different scale of the color bar) and the smaller the cross-section the smaller the mass splitting should be in order to produce a nuclear recoil. The marginal posterior pdf is again the only one which is unimodal and for which we can quote a best fit point: $m_{\rm DM} = 7.7$ GeV, $\sigma_n^{SI}= 4 \times 10^{-40}{\rm cm^2}$ and $\delta = 6.1$ keV. 
\item CRESST: inelastic SI interactions fit the data in a wide range of masses and cross-sections. All values of $\delta$ are allowed, as can be seen from the right panel figure~\ref{fig:deltaSMH}. 
\item KIMS: the exclusion bound is less constraining than the one quoted by the collaboration as a consequence of the marginalisation over the quenching factors and $\alpha$ background. 
\end{itemize}

\paragraph*{Isospin violating SI scattering} 
The comments below refer to figure~\ref{fig:All_NFW} (right panel) and figure~\ref{fig:fnfpSMH}, and are valid both for SMH and marginalised astrophysical case.

\begin{itemize}
\item DAMA: again two regions are defined, due to the multi-target detector, one at small DM masses and one for masses $\sim 100$ GeV. Both regions denote the same trend with respect to $f_n/f_p$: the smaller the cross-section is, the more negative the $f_n/f_p$ value becomes, as shown by the correlation between $m_{\rm DM}$, $\sigma^{SI}_n$ and $f_n/f_p$ in figure~\ref{fig:fnfpSMH} (left panel). 
\item CoGeNT: the detection region has a similar dependence on $f_n/f_p$ as the DAMA one (central panel figure~\ref{fig:fnfpSMH}). The values that maximize the unimodal posterior pdf are $m_{\rm DM} = 7.5$ GeV, $\sigma_n^{SI}= 2 \times 10^{-40}{\rm cm^2}$ and $f_n/f_p = 0.6$. 
\item CRESST: the excess can be explained by a wide range of masses and cross-section values and for all possible values of $f_n/f_p$ (right panel in figure~\ref{fig:fnfpSMH}).  
\item Exclusion bounds:  SIMPLE, KIMS and CDMSSi are less restrictive for this physical scenario and do not show particular features in their nuisance parameters. 
\end{itemize}

\bibliographystyle{apsrev4-1}
\bibliography{biblio}

\end{document}